\documentclass[twocolumn,aps,prb,floatfix,longbibliography,superscriptaddress]{revtex4-2}
\usepackage[utf8]{inputenc}
\usepackage[T1]{fontenc}
\usepackage{amsmath, amssymb, amsfonts}
\usepackage{graphicx}
\usepackage{hyperref}
\usepackage{physics}
\usepackage{xcolor}
\usepackage{microtype}
\usepackage{bm}

\definecolor{linkcolor}{rgb}{0,0,0.5}
\hypersetup{
    colorlinks=true,
    linkcolor=linkcolor,
    citecolor=linkcolor,
    urlcolor=linkcolor
}

\begin{document}

\title{\textbf{Frustration-induced degenerate spin state with \textit{up-up-down-down} ordering in corner-connected Heisenberg square-plaquettes}}

\author{Pritam Manna}
\affiliation{Solid State Physics Division, Bhabha Atomic Research Centre, Mumbai 400085, India}

\author{A.~K.~Bera}
\email{akbera@barc.gov.in}
\affiliation{Solid State Physics Division, Bhabha Atomic Research Centre, Mumbai 400085, India}
\affiliation{Homi Bhabha National Institute, Anushaktinagar, Mumbai 400094, India}

\date{\today}

\begin{abstract}
We investigate frustrated magnetism of a corner-connected square plaquette Heisenberg model with exchange interactions along the edges $(J_1)$, along inter-plaquette links $(J_2)$, and along square diagonals $(J_3)$. Using Luttinger--Tisza (LT) minimization of the Fourier interaction matrix $\mathcal J(\mathbf q)$ together with large-scale Monte Carlo (MC) simulations, we obtain a classical low-temperature magnetic phase diagram in the normalized plane $(J_1/|J_2|, J_3/|J_2|)$. The two methods play complementary roles: the LT analysis provides the zero-temperature candidate ordering wave vectors, while the MC simulations elucidate the resulting ordering tendencies under the hard-spin constraint at low but finite temperatures. Three regimes emerge at low temperatures, an antiferromagnetic phase (AF), a ferromagnetic phase (FM) and a degenerate spin state with \textit{up-up-down-down (uudd)} ordering (DS). For the frustrated degenerate spinstate, LT exhibits line-like minima along $q_x=\pm q_y$ in the $hk$-plane, revealing a highly degenerate spin configuration which violates the hard-spin constraint. The MC results uncover a state with quasi-two-dimensional ordering. The DS regime is intrinsically multi-$\mathbf{q}$: the ordered texture assembles itself from symmetry-related modes on the lines, producing a ``\textit{uudd}'' spin arrangement comprising of distinctive strong antiferromagnetic correlations on diagonals, ferromagnetic correlations on corner links, and highly suppressed correlations on the edges. The field-temperature phase diagram for a representative parameter point in the DS regime, determined by MC simulations, yields field induced distinct regions of negatively and positively correlated layers. These two regions are separated by a curve corresponding to negligible inter-layer correlations. Our LT+MC framework delivers a well-controlled classical baseline and operational diagnostics for corner-connected frustrated square plaquette systems, thereby benchmarking future approaches that incorporate quantum-fluctuation effects.
\end{abstract}

\maketitle

\section{Introduction}\label{sec:intro}

Magnetic frustration arises from competing interactions that cannot be simultaneously satisfied giving rise to complex magnetic ground states and unconventional collective behavior \cite{Balents2010Nature,MoessnerRamirez2006PT,LacroixMendelsMila2011}. In classical spin systems, such competition often produces extensively degenerate manifolds and nontrivial magnetic order, including spiral orders and multi-$\mathbf{q}$ textures \cite{Villain1980JPhys,Henley1989PRL,Okubo2012PRL,Liu2024PRB}. The isotropic Heisenberg model with multiple competing exchange interaction is a natural arena to explore these effects. 

Much of the canonical literature has emphasized geometries built from corner-sharing triangles or tetrahedra, kagome, pyrochlore, and hyperkagome \cite{Balents2010Nature,GardnerGingrasGreedan2010RMP,Okamoto2007PRL}. By contrast, corner-connected square plaquettes provide a complementary route to frustration and remains unexplored. The competition between exchange interactions can be neatly studied in the momentum space \cite{LuttingerTisza1946}: nearest-neighbor couplings along the square edges, diagonal bonds within the square, and inter-square links (joining squares at connected corners) [Fig. 1(a)] conspire to produce competing minima in the lowest eigenvalue of $J(\mathbf{q})$ \cite{Rastelli1979Physica,Sindzingre2010JPCS}. This viewpoint naturally connects to the well-studied $J_1$--$J_2$–$J_3$ Heisenberg square lattice model (where plaquettes are connected by sharing their edges), which yields collinear, ferromagnetic, and incommensurate spiral phases in two dimensions \cite{Rastelli1979Physica,Sindzingre2010JPCS, PhysRevB.93.085132, PhysRevB.84.064407, PhysRevB.82.094412, PhysRevB.88.104401}. In contrast, the ground state phase diagram of corner connected square plaquettes remains unexplored. 

Corner-connected square plaquette based frameworks realized in the rare-earth series $\mathrm{\textit{X}}_2\mathrm{Re}_3\mathrm{Si}_5$ (\textit{X} a rare-earth element) \cite{Sharma2022PRM} are of growing interest. In these materials, the magnetic atoms are organized in a network of corner-connected square-plaquettes to form a three dimensional lattice [Fig. 1(c)], thereby enabling direct tuning of inter-square couplings through connectivity rather than through pure planar geometry. This geometrical difference motivates a systematic exploration of the magnetic behavior as a function of the relative exchange interaction strengths along square edges ($J_1$), inter-square corner connecting links ($J_2$), and square diagonals ($J_3$), and invites comparison with both square-lattice $J_1$–$J_2$–$J_3$ models and three-dimensional frustrated frameworks \cite{Rastelli1979Physica,Sindzingre2010JPCS}.

In this work, we investigate the magnetic behavior and the magnetic phase diagram of the classical Heisenberg model on a frustrated lattice comprising of a three-dimensional network of corner-connected square units stacked periodically along the crystallographic $c$ axis. We fix the inter-square coupling $J_2$ to be ferromagnetic and use it to set the energy scale. Within this convention, the normalized parameter space $(J_1/|J_2|, J_3/|J_2|)$ captures the essential competition between the exchange interactions. Applied to this geometry, momentum-space minimization using the Luttinger–Tisza (LT) approach provides a transparent organizing principle for the classical ground states because the minimal eigenmode of $J(\mathbf{q})$ directly encodes the candidate ordering wave vector(s) \cite{LuttingerTisza1946,LyonsKaplan1960}. The LT analysis enumerates three robust regimes across the parameter plane: a collinear antiferromagnetic (AF) state with alternating spins on the square units and parallel spins along the corner connecting bonds, a ferromagnetic (FM) phase, and a highly degenerate multi-\textit{q} magnetic phase characterized by extended manifolds of nearly equivalent wave vectors. While LT is exact for Bravais lattices with a single spin per unit cell and often remains remarkably predictive for multi-sublattice problems, it can artificially enhance degeneracies by relaxing the local spin-length constraint \cite{LyonsKaplan1960,Henley2010ARCMP}. Therefore, a stringent test of the LT topology requires explicit enforcement of the constraints and the inclusion of thermal fluctuations. Classical Monte Carlo (MC) simulations serve this role: by sampling spin configurations at low (and finite) temperatures with periodic boundary conditions, one can reconstruct the phase diagram from two-point spin-spin correlation functions and determine various order parameters, response functions, and static spin structure factors, thus locating phase boundaries \cite{LandauBinder2021,NewmanBarkema1999,Metropolis1953}.
An applied magnetic field is an essential control parameter for frustrated magnets. Because the Zeeman term couples linearly to the uniform magnetization, it can lift degeneracies within the spiral manifold, induce conical or fan states, and stabilize multi-$\mathbf{q}$ superpositions in the presence of weak anisotropies \cite{Nagamiya1967SSP,Okubo2012PRL}. To elucidate these effects in the corner-connected–square network, we augment the low-temperature magnetic phase diagram with field–temperature ($H$–$T$) maps at two representative points in the parameter-space: one in the highly degenerate regime with \textit{uudd} ordering and one in the collinear AF region. We construct these $H$-$T$ phase diagrams by tracking correlation-based indicators and thermodynamic observables across $(H,T)$ at fixed $(J_{1}/|J_{2}|, J_{3}/|J_{2}|)$, thereby delineating field and temperature induced transitions.

In short, our contributions are twofold. First, we establish a comprehensive classical baseline for the corner-connected square network by mapping its low-temperature phase diagram across $(J_1/|J_2|, J_3/|J_2|)$ and delineating AF, FM, and DS regimes using a combination of LT reasoning and hard-spin constraint enforcing MC simulations. The DS state shows an emergent quasi-two-dimensional correlation structure, constituted by strongly correlated $uudd$ spin-order along diagonal planes and weak correlations outside the plane. The observed DS phase is distinct and has no direct analogue in other well-studied frustrated lattices such as kagome, square, or honeycomb networks  The observed DS phase is distinct and has no direct analogue in other well-studied frustrated lattices such as kagome, square, or honeycomb networks \cite{ChalkerHoldsworthShender1992PRL,ReimersBerlinsky1993PRB,Mulder2010PRB,PhysRevB.93.085132}. Second, we construct $H$–$T$ phase diagrams at selected couplings, thereby exposing the differences in field-induced magnetic responses between the DS and AF regions. In the DS regime, the $H-T$ phase diagram reveals field induced distinct regions with negatively and positively correlated layers separated by a curve corresponding to a stack of layers between which correlations are highly suppressed. These results furnish benchmarks for future calculations (e.g., spin-wave theory, exact diagonalization, tensor-network approaches, etc.) \cite{Auerbach1994,White1992PRL,Orus2014AOP}.

\section{Model and Methods}
\subsection{Model}
We consider a frustrated Heisenberg spin model constructed from corner–connected square plaquettes stacked periodically along the crystallographic $c$ axis [Figs.~1(a--c)]. Within each plaquette, the nearest–neighbor exchange interaction $J_{1}$ couples spins located at adjacent corners along the edges of the square. In addition to these edge couplings, the diagonal bonds inside each square introduce further in–plane interactions, parameterized by $J_{3}$. The latter term introduces magnetic frustration by competing with the nearest–neighbor exchange interaction $(J_1)$. In the present study, we adopt the following sign convention for the exchange couplings: a positive exchange constant corresponds to an antiferromagnetic interaction, favoring an antiparallel alignment of neighboring spins, whereas a negative exchange constant corresponds to a ferromagnetic interaction, favoring a parallel spin alignment.

\begin{figure*}
    \centering
     \includegraphics[width=\linewidth]{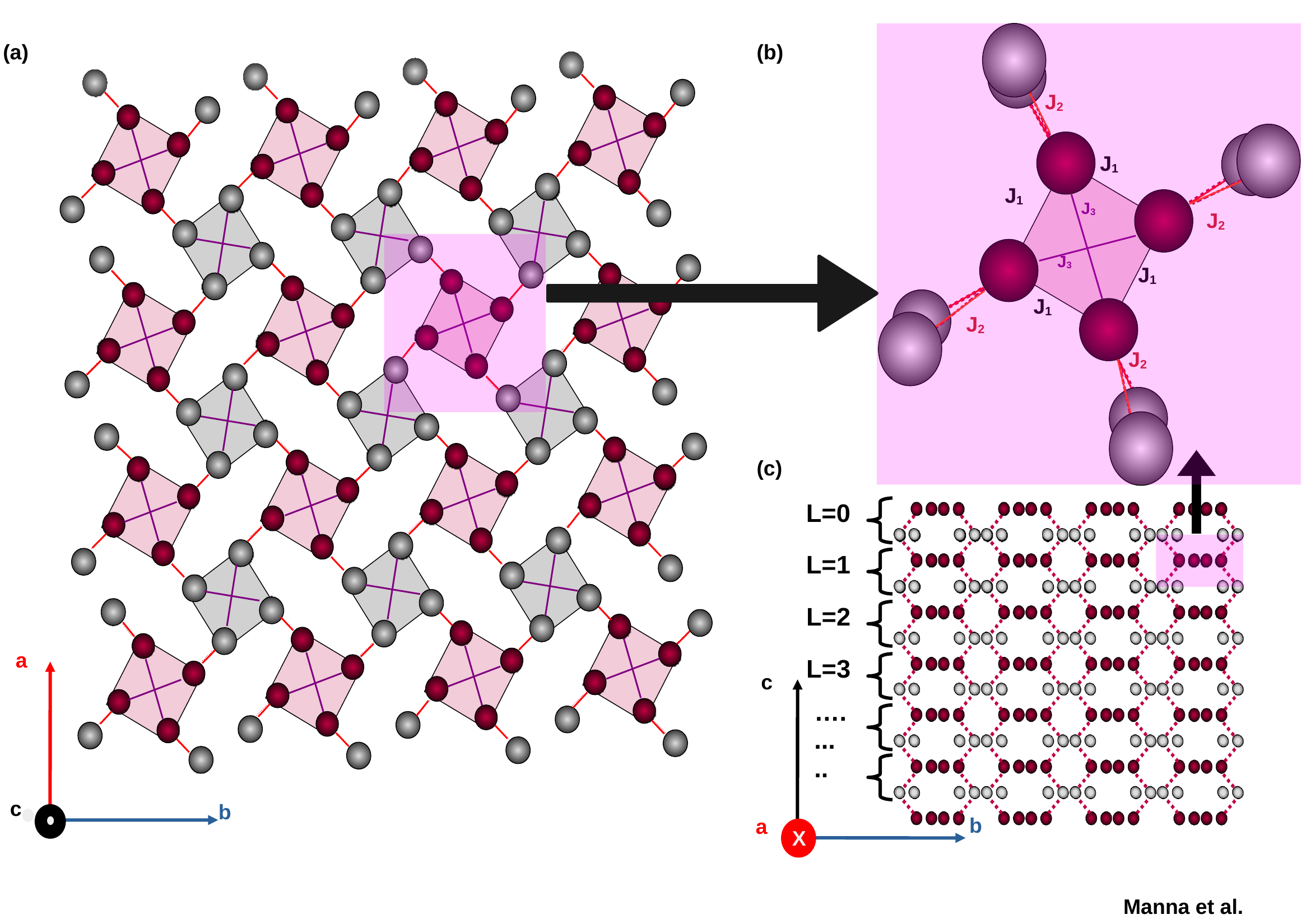}
    \caption{
\textbf{(a)} Top view of the corner-connected square network, showing the arrangement of square plaquettes in the basal plane. The pink and gray squares are located in the adjacent layers along the \textit{c}-axis.
\textbf{(b)} Top view of the conventional unit cell (highlighted in magenta) indicating exchange paths: nearest-neighbor \(J_{1}\) along plaquette edges, diagonal \(J_{3}\) within each plaquette, and inter-plaquette \(J_{2}\) linking neighboring plaquettes (located on adjacent layers) through the corners.
\textbf{(c)} Side view along the \(c\) axis, highlighting the three-dimensional stacking and connectivity of the network. For representation purposes, the layers are marked with numbers: $L$=0,1,2,3,... etc . 
}

    \label{fig:Fig 1}
\end{figure*}

The square plaquettes are connected through their corners $(J_2$ bonds$)$, thereby forming a network of corner–connected squares stacked along the $c$ axis [Fig. 1(c)]. In the present work, we focus on the regime where $J_{2}$ is ferromagnetic, thereby stabilizing parallel alignment of spins across successive layers of plaquettes. This restriction reduces the parameter space and allows us to highlight the competition between the edge exchange $J_{1}$ and the diagonal interaction $J_{3}$.

The resulting Hamiltonian can be written as
\begin{equation}
\begin{split}
\mathcal{H} = \,
& J_{1}\sum_{\langle i,j\rangle \in \hspace{0.2mm} edges} \mathbf{S}_{i}\cdot\mathbf{S}_{j}
+ J_{3}\sum_{\langle\langle i,j \rangle\rangle \in \hspace{0.2mm} diag} \mathbf{S}_{i}\cdot\mathbf{S}_{j} \\
& + J_{2}\sum_{\langle i,k \rangle  \in \hspace{0.2mm} corners} \mathbf{S}_{i}\cdot\mathbf{S}_{k}
- h_z \sum_i S^{z}_i ,
\end{split}
\label{eq:Hamiltonian}
\end{equation}
where $\mathbf{S}_{i}$ denotes the spin at site $i$. The first summation runs over all nearest–neighbor pairs along the sides of each square, the second covers the diagonal next–nearest neighbors within the square plaquettes, the third accounts for inter–plaquette couplings along the corner connecting bonds between the nearest square units and the last term corresponds to the Zeeman energy due to an externally applied magnetic field along the $c$ direction. This geometry of corner–connected square plaquettes stacked along the $c$ axis provides a natural extension beyond both the simple square lattice and layered honeycomb networks studied previously, offering a fertile ground for the emergence of unconventional magnetic ground states under competing interactions.

Our analysis proceeds in two stages. First, we determine the classical magnetic ground states using the Luttinger–Tisza (LT) method, which minimizes the Fourier–transformed exchange energy over ordering vectors $\mathbf{q}$ in the Brillouin zone (BZ) while relaxing the local hard–spin constraint to a global norm condition \cite{LuttingerTisza1946,LyonsKaplan1960}. This yields candidate single–$\mathbf{q}$ configurations and a rigorous lower bound to the total classical exchange energy. The resulting spin configurations have subsequently been checked against the exact hard–spin constraint $|\mathbf{S}_i|=1$ for every sublattice site; if the lowest–energy LT eigenmode cannot be normalized on all sublattices, the corresponding solution is unphysical for the Heisenberg model and typically signals either multi–$\mathbf{q}$ order or an extensively degenerate manifold of low–lying states in momentum space \cite{LyonsKaplan1960,Henley2010ARCMP}. The second stage employs large–scale classical Monte Carlo (MC) simulations with parallel tempering to validate LT candidates as well as predict the magnetic phase diagram of the system as a function of temperature and magnetic fields \cite{Metropolis1953,LandauBinder2021,HukushimaNemoto1996,Geyer1991}.

\subsection{Luttinger Tisza Method}
The conventional unit cell of the studied corner connected square plaquettes system contains eight magnetic sites [Fig.~1(b)] at fractional coordinates $\boldsymbol{\rho}_\mu=(\rho_\mu^x,\rho_\mu^y,\rho_\mu^z)$, $\mu=1,\dots,8$ (Table I). We denote the Bravais translation between two conventional cells by $\mathbf{R}$ and the exchange coupling between the $\mu^{th}$ site in the origin cell and the $\nu^{th}$ site in the cell at $\mathbf{R}$ by $J_{\mu\nu}(\mathbf{R})$. For $\mathbf{R}=0$, the sites are taken within the conventional unit cell. Magnetic self–interaction is unphysical, hence the diagonal terms of the interaction matrix vanish (at $ \mathbf{R} = 0$). The Bloch–transformed interaction matrix in momentum space is\begin{equation}
\mathcal{J}_{\mu\nu}(\mathbf{q})=\sum_{\mathbf{R}} J_{\mu\nu}(\mathbf{R})\;
e^{\,i\,\mathbf{q}\cdot\mathbf{R}},
\end{equation}
and the LT energy is given by the smallest eigenvalue $\lambda_{\min}(\mathbf{q})$ of the $8\times8$ Hermitian matrix $\mathbf{\mathcal{J_{\mu\nu}}}(\mathbf{q})$ (Equation 15).

We discretize the first BZ ($BZ_1$) on a uniform grid and diagonalize $\mathcal{J_{\mu\nu}}(\mathbf{q})$ at each mesh point to obtain the eigenbands $\{\lambda_n(\mathbf{q})\}_{n=1}^{8}$. The ordering vector $\mathbf{q}_\star$ is identified by minimizing the lowest branch in the first Brillouin zone,
\begin{equation}
\mathbf{q}_\star=\underset{\mathbf{q}\in BZ_1}{\arg\min}\;\lambda_{\min}(\mathbf{q})\,.
\end{equation}

For the isotropic Heisenberg case, minimizing the lowest band of 
$\mathcal J_{\mu\nu}(\mathbf q)$ yields a normalized eigenvector 
$v(\mathbf q_{\star})=(v_1,\dots,v_M)\in\mathbb C^M$ (here $M{=}8$) whose entries furnish the complex sublattice amplitudes for a single–$\mathbf q$ state.  A single–$\mathbf q$ LT state satisfies the hard–spin constraint  $\lvert \mathbf S_i(\mathbf R)\rvert=1$ only if the moduli of the eigenvector components are sublattice–independent,
\[
\lvert v_i(\mathbf q_{\star})\rvert = \text{const.}\quad\forall\,i,
\]
so that there exist orthonormal spin–space vectors 
$\hat{\mathbf u},\hat{\mathbf v}$ and phases $\phi_i$ (obtained from the components of $v(\mathbf q_{\star})$) with
\[
\mathbf S_\mu(\mathbf R)
=\cos\!\big(\mathbf q_{\star}\!\cdot\!\mathbf R+\phi_\mu\big)\,\hat{\mathbf u}
+\sin\!\big(\mathbf q_{\star}\!\cdot\!\mathbf R+\phi_\mu\big)\,\hat{\mathbf v}.
\]
(For a Bravais lattice, $M{=}1$, this condition is automatic \cite{LyonsKaplan1960}).
If this modulus–equality condition fails, the LT minimum only provides a
\emph{lower bound} to the classical ground–state energy, and the actual ground
state may be constructed from a superposition of modes drawn from the
degenerate LT–minimum manifold 
$\mathcal M=\{\mathbf q:\lambda_{\min}(\mathbf q)=\lambda_{\min}\}$ in order to enforce
$\lvert \mathbf S_\mu(\mathbf R)\rvert=1$ on every site. 

 We follow the standard LT construction for multi–sublattice Heisenberg models, diagonalizing the $8\times8$ interaction matrix $\mathcal{J_{\mu\nu}}(\mathbf{q})$ on a uniform Brillouin–zone mesh and identifying $\mathbf{q}_{\star}$ by minimizing $\lambda_{\min}(\mathbf{q})$ \cite{LuttingerTisza1946,LyonsKaplan1960}.

\subsection{Monte Carlo Method}
Because LT relaxes the local spin–length constraint and therefore cannot resolve regions with extended $\mathbf{q}$–space degeneracy or potential multi–$\mathbf{q}$ order, we complement it with large–scale classical MC simulations. We first map the bond-resolved correlation functions [Eq. 8] at low temperatures $(T = 0.1 |J_2|)$ in the $(J_{3}/|J_{2}|,\,J_{1}/|J_{2}|)$ plane for lattice sizes of $N\times N\times N$ ($N{=}3,7$) supercells (having eight spins per unit cells, thus having 216 and 2744 spin sites respectively). This allows us to construct a field-temperature phase diagram at representative points in the parameter space. For each parameter pair $(J_{3}/|J_{2}|,\,J_{1}/|J_{2}|)$, we perform Monte Carlo simulations using single–spin Metropolis algorithm \cite{Metropolis1953,LandauBinder2021}. We equilibrate and measure two–point correlators and other order parameters, which together characterize the dominant low-temperature spin correlations and allow a comparison with the soft-mode structure obtained from LT analysis. We perform our calculations using $2\times10^{6}$ Monte Carlo steps, with the first half discarded for thermal equilibration, to compute thermodynamic order parameters and bond-resolved correlation functions. One Monte Carlo sweep (MCS) consists of $8N^3$ attempted updates; a trial rotation at a randomly chosen site is accepted with probability
\begin{equation}
P_{\mathrm{acc}}=\min\!\left[1,\exp(-\Delta E/T)\right].
\end{equation}
Given the frustrated character of the model, we employ parallel tempering (replica–exchange MC) \cite{HukushimaNemoto1996,Geyer1991,Katzgraber2006,Kofke2002} typically with 24–72 replicas on a temperature ladder. After every 12 MCS we attempt exchanges between neighboring replicas, accepting swaps with probability
\begin{equation}
P_{\mathrm{swap}}=\min\!\left[1,\exp\!\bigl((\beta_{i}-\beta_{j})(E_{i}-E_{j})\bigr)\right],
\end{equation}
Thermodynamic observables and their probability distributions, obtained over the $(J_1,J_3)$ plane and temperature grid, serve as primary diagnostics for characterizing phases and thermal transitions. The energy distribution $P(E,T,J_1/|J_2|,J_3/|J_2|)$ is the normalized histogram of energy per spin at a fixed temperature. Its evolution with $T$ is a sensitive probe: a single–peaked Gaussian profile is typical of a continuous transition (second order), while the emergence of two well–separated peaks signals a first–order transition and presence of a latent heat \cite{ChallaLandauBinder1986,LeeKosterlitz1990,LeeKosterlitz1991}. 

Similarly, the distributions of relative spin angles,
\begin{equation}
    P(\bm{\Theta}_{1,2,3}) = P\big( \angle(\mathbf{S}_i,\mathbf{S}_j) \big)_{(i,j)\in J_{1,2,3}},
\end{equation}
measured along $J_1$ edges, $J_2$ corner–link bonds, and $J_3$ diagonals, reveal the relative arrangement of spins in the conventional unit cell. The total magnetization per spin,
\begin{equation}
    M = \frac{1}{N}\Big\langle \Big|\sum_{\mathbf{R},\alpha} \mathbf{S}_{\mathbf{R},\alpha}\Big| \Big\rangle,
\end{equation}
acts as a global measure of ferromagnetic order. 
To resolve local ordering tendencies, we compute the bond–resolved spin–spin correlations
\begin{equation}
    X_2(T,J_n) = \langle \mathbf{S}_i \cdot \mathbf{S}_j \rangle_{(i,j)\in J_n}, \quad n=1,2,3,
\end{equation}
which quantify the average nearest–neighbor alignment along $J_1$, $J_2$, and $J_3$ bonds. In addition to these bond-resolved diagnostics, we have also calculated full-lattice reference-site spin-spin correlation maps for representative low-temperature configurations to verify whether the local correlation rules inferred from the three exchange paths propagate over the full lattice. In these maps, a selected spin is chosen as the reference point, and its scalar correlations with all other spins in the simulated lattice are evaluated. The analysis was performed for several symmetry-related reference points. 

In the antiferromagnetic sector, we define the plaquette–adapted staggered magnetization
\begin{equation}
    \mathbf{M}_s = \frac{1}{N}\sum_{\mathbf{R}} \sum_{\alpha=1}^{8} \eta_\alpha \, \mathbf{S}_{\mathbf{R},\alpha},
\end{equation}
where $\boldsymbol{\eta}=(+1,-1,+1,-1,-1,+1,-1,+1)$ encodes alternating signs within each plaquette and parallel alignment across corner links. A nonzero plaquette adapted staggered magnetization $(\mathbf{M}_s)$ indicates a long–range AF order. 

In the DS regime, the low-temperature  MC simulations show strong $uudd$-type correlations develop along the diagonal-corner link trajectories, while, weak correlations through the square edges. To identify this, we employ the four–spin order parameter $(O _2)$. 
\begin{multline}\label{eq:O2_split}
O_2=\frac{1}{N_{\mathrm{paths}}}\sum_{\mathrm{paths}(D)}
\biggl|\frac{1}{N_D N_L}
\sum_{\substack{L\in \mathrm{Layers}\\ (a,b)\in D}}
\tfrac14\bigl(\mathbf{S}_{c(a),L}+\mathbf{S}_{a,L}\\
-\mathbf{S}_{b,L}-\mathbf{S}_{c(b),L}\bigr)\biggr|.
\end{multline}

which captures the characteristic $uudd$ motif along diagonally connected conventional cells and is closely related to diagnostics used in recent multiple–$\mathbf{q}$ literature \cite{Okubo2012PRL,Hayami2024PRB,Gutzeit2022NatCommun}. Here, the inner sum is taken over a particular diagonal trajectory (\textbf{D}). The spins $\mathbf{S_{a,L}}$, ${ \mathbf{S_{b,L}}}$ are chosen along the diagonal bond in the square plaquette (at layer \textbf{L}). The spins $\mathbf{S_{c(a),L}}$, ${ \mathbf{S_{c(b),L}}}$  represent nearest to  $\mathbf{S_{a,L}}$, ${ \mathbf{S_{b,L}}}$ along the corner connecting bonds (at layer \textbf{L}) [Fig. 6(a)]. The layers are ferromagnetically coupled, hence, the same structure repeats along the \textit{c} axis. $N_D$ represents the number of sites along a diagonal trajectory \textbf{D}. $N_L$ represents the number of layers stacked along the c-axis. $N_{paths}$ represents the number of independent diagonal trajectories. The outer sum averages over the diagonal trajectories used to characterize the $uudd$-type correlation pattern. Thermodynamic response functions complement these order parameters. The specific heat per spin, exhibits sharp peaks where entropy is released. The uniform susceptibility$(\chi)$, staggered susceptibility$(\chi_s)$ and the susceptibility in the order parameter $O_2$ $(\chi_{_{O_2}})$ are given below,
\begin{equation}
\chi_{X}=\frac{N}{T}\big(\langle X^{2}\rangle-\langle X\rangle^{2}\big),\qquad X\in{M,M_s,O_2},
\end{equation}
Here $X$ denotes a generalized order parameter. The specific heat ($C$) is defined:
\begin{equation}
    C = \frac{N}{T^2} \big( \langle E^2 \rangle - \langle E \rangle^2 \big),
\end{equation}
These track the growth of phase specific correlations, diverging at the corresponding critical temperatures of the FM $(\chi)$, AF $(\chi_s)$ and DS $(\chi_{_{O_{2}}})$ phases. The static spin structure factor
\begin{equation}
    S(\mathbf{q}) = \frac{1}{N}\sum_{i,j} \langle \mathbf{S}_i \cdot \mathbf{S}_j \rangle e^{i \mathbf{q}\cdot (\mathbf{r}_i - \mathbf{r}_j)},
\end{equation}
reveals the propagation vector via Bragg peaks and enables direct comparison with LT calculations and neutron–scattering conventions \cite{Squires2012,Lynn2012MagNeutron}. 

\begin{table}
\caption{Fractional coordinates $(\rho_x,\rho_y,\rho_z)$ of Heisenberg spins in the conventional unit cell.}
\label{tab:Ho_frac}
\begin{ruledtabular}
\begin{tabular}{lccc}
$\mu$& $\rho^{x}_\mu$& $\rho^{y}_\mu$& $\rho^{z}_\mu$\\
\hline
1& 0.5704& 0.2606& 0.0\\
2& 0.7394& 0.5704& 0.0\\
3& 0.4296& 0.7394& 0.0\\
4& 0.2606& 0.4296& 0.0\\
5& 0.7606& 0.0704& 0.5\\
6& 0.0704& 0.2394& 0.5\\
7& 0.2394& 0.9296& 0.5\\
8& 0.9296& 0.7606& 0.5\\
\end{tabular}
\end{ruledtabular}
\end{table}
Finally, to characterize field responses, we select representative points in the AF and  DS regimes and perform MC on larger lattices, $11\times11\times11$ (10\,648 spins). The phase boundaries are inferred from peaks in $C(T)$, $\chi(T)$, $\chi_{s}(T)$, two–point spin–spin correlations, the $O_2$ parameter, and the evolution of $S(\mathbf{q})$. In fields, degeneracy lifting can stabilize conical/fan states and multi–$\mathbf{q}$ superpositions \cite{Okubo2012PRL,Hayami2024PRB}.

\section{Results and Discussion}
\subsection{Luttinger--Tisza (LT) Analysis}
\begin{figure*}
    \centering
    \includegraphics[width=\textwidth]{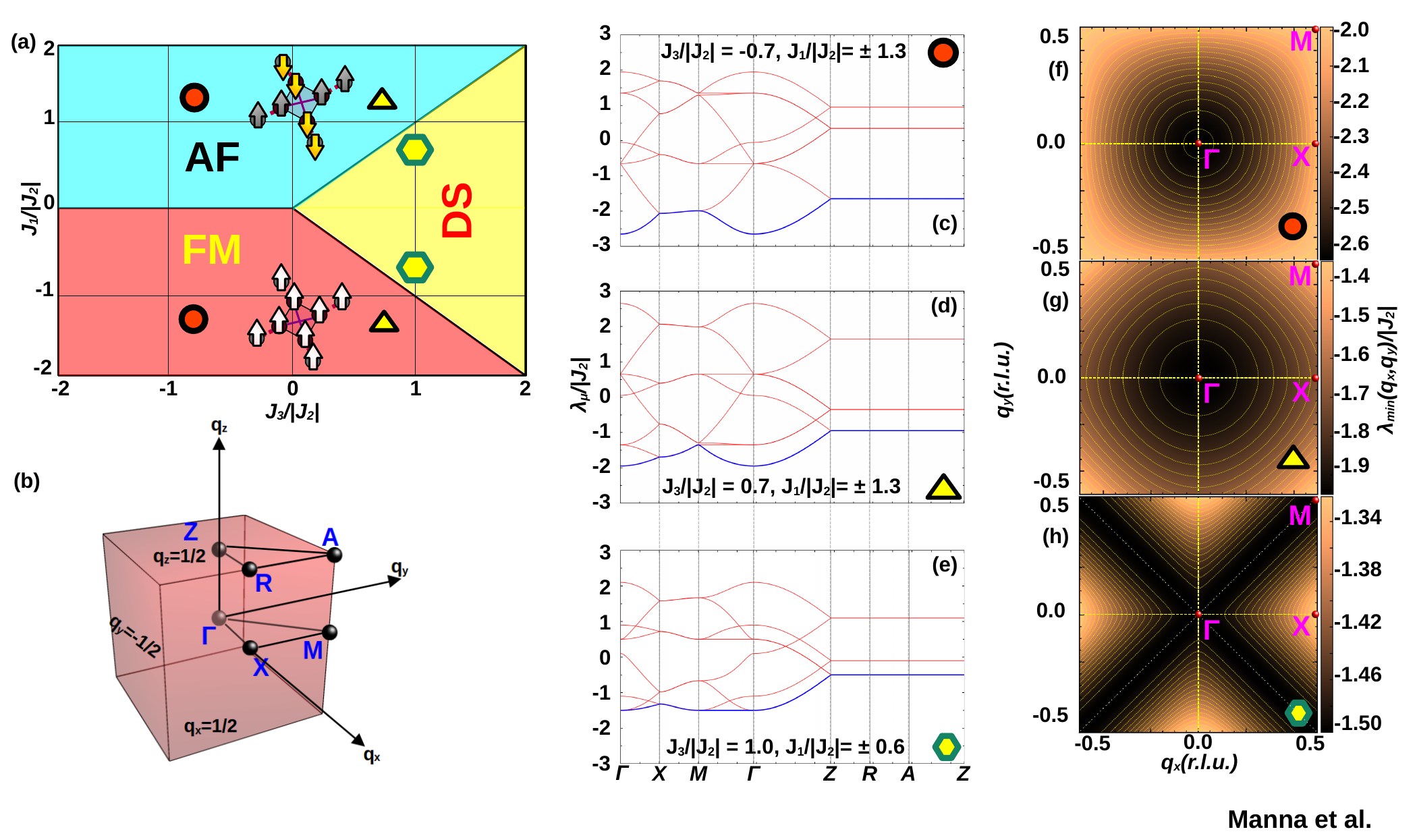}
    \caption{
    (a) Classical ground-state magnetic phase diagram obtained from a Luttinger--Tisza (LT) analysis; Representative parameter points used for the band-structure calculations are indicated on the diagram by symbols:  a circle, a triangle, and a hexagon. Circle: $J_3/|J_2|=-0.7$, $J_1/|J_2|=\pm 1.3$ , Triangle: $J_3/|J_2|=0.7$, $J_1/|J_2|=\pm 1.3$, Hexagon: $J_3/|J_2|=1.0$, $J_1/|J_2|=\pm 0.6$  
    (b) High-symmetry path in the first Brillouin zone (of the conventional unit cell) adopted for the interaction-matrix band-structure plots, $\Gamma\!-\!X\!-\!M\!-\!\Gamma\!-\!Z\!-\!R\!-\!A\!-\!Z$.
    (c--e) Eigenvalue dispersions ($\lambda_{\mu}/|J_2|, \mu=1,2,3,4,5,6,7,8$) of the LT interaction matrix $\mathcal{J}(\mathbf{q})$ along the path in (b) for the representative points in (a); the lowest branch sets the classical LT energy $E_{\mathrm{LT}}(\mathbf{q})$ and identifies candidate ordering vectors.
    (f--h) Heat maps of the minimum eigenvalue $\lambda_{\min}(\mathbf{q})$ (in units of $|J_2|$) on the $q_z\!=\!0$ plane for the representative points on the ground-state magnetic phase diagram (a).}
    \label{fig:Fig 2}
\end{figure*}
To determine the magnetic ground state of the model described in Eq. 1, we use the Luttinger--Tisza method\cite{LuttingerTisza1946,LyonsKaplan1960}. We build a Bloch interaction matrix and diagonalize it for each $\mathbf{q}$ in the first Brillouin zone. We work in reciprocal--lattice units (r.l.u.), so $q_x a\!\to\!2\pi q_x$ (and similarly for $q_y,q_z$), and adopt a gauge in which basis phases (for the eight sub-lattices) are absorbed into the eigenvectors. Writing the Bravais translation as $\mathbf R=(n_x,n_y,n_z)$ in units of lattice parameters: $a,b,c$, the isotropic Heisenberg interaction matrix in momentum space becomes:
\vspace{0mm}
\begin{equation}
\mathcal J_{\mu\nu}(\mathbf q)
=\sum_{(n_x,n_y,n_z)\in\mathbb{Z}^3} J_{\mu\nu}(\mathbf R)\,
e^{\,i\,2\pi(n_x q_x+n_y q_y+n_z q_z)}
\end{equation} 
For the studied spin system with $J_1$ (exchange interactions along a square edge), $J_3$ (along the square diagonals) and $J_2$ (along the bonds connecting two adjacent square corners) and a sublattice ordering given in Table I, the interaction matrix is in momentum space is given in Eq. 15. Here, $\alpha = \exp(i2\pi q_x)$, $\beta = \exp(i2\pi q_y)$ and $\gamma = \exp(i2\pi q_z)$. $(q_x,q_y,q_z)$ are measured in reciprocal lattice units (r.l.u.). We obtain the ordering vector $\mathbf{q_{\star}}$ by minimizing the lowest eigenvalue $\lambda_{min}(\mathbf{q})$ of $\mathcal{J}(\mathbf{q})$ over the first Brillouin zone. The corresponding normalized eigenvector furnishes the sublattice amplitudes/phases.
\begin{widetext}
\begin{equation}
\setlength{\arraycolsep}{4pt}\renewcommand{\arraystretch}{0.95}
\mathcal J(\mathbf q)=
\begin{pmatrix}
0 & J_1\alpha^{-1} & J_3(\alpha\beta)^{-1} & J_1\beta^{-1} & 0 & 0 & 0 & J_2(1+\gamma) \\
J_1\alpha & 0 & J_1\beta^{-1} & J_3\alpha\beta^{-1} & J_2(1+\gamma) & 0 & 0 & 0 \\
J_3\alpha\beta & J_1\beta & 0 & J_1\alpha & 0 & J_2(1+\gamma) & 0 & 0 \\
J_1\beta & J_3\alpha^{-1}\beta & J_1\alpha^{-1} & 0 & 0 & 0 & J_2(1+\gamma) & 0 \\
0 & J_2(1+\gamma^{-1}) & 0 & 0 & 0 & J_1 & J_3 & J_1 \\
0 & 0 & J_2(1+\gamma^{-1}) & 0 & J_1 & 0 & J_1 & J_3 \\
0 & 0 & 0 & J_2(1+\gamma^{-1}) & J_3 & J_1 & 0 & J_1 \\
J_2(1+\gamma^{-1}) & 0 & 0 & 0 & J_1 & J_3 & J_1 & 0
\end{pmatrix}
\label{eq:Jq_explicit}
\end{equation}
\end{widetext}

\begin{figure*}
    \centering
    \includegraphics[width=\linewidth]{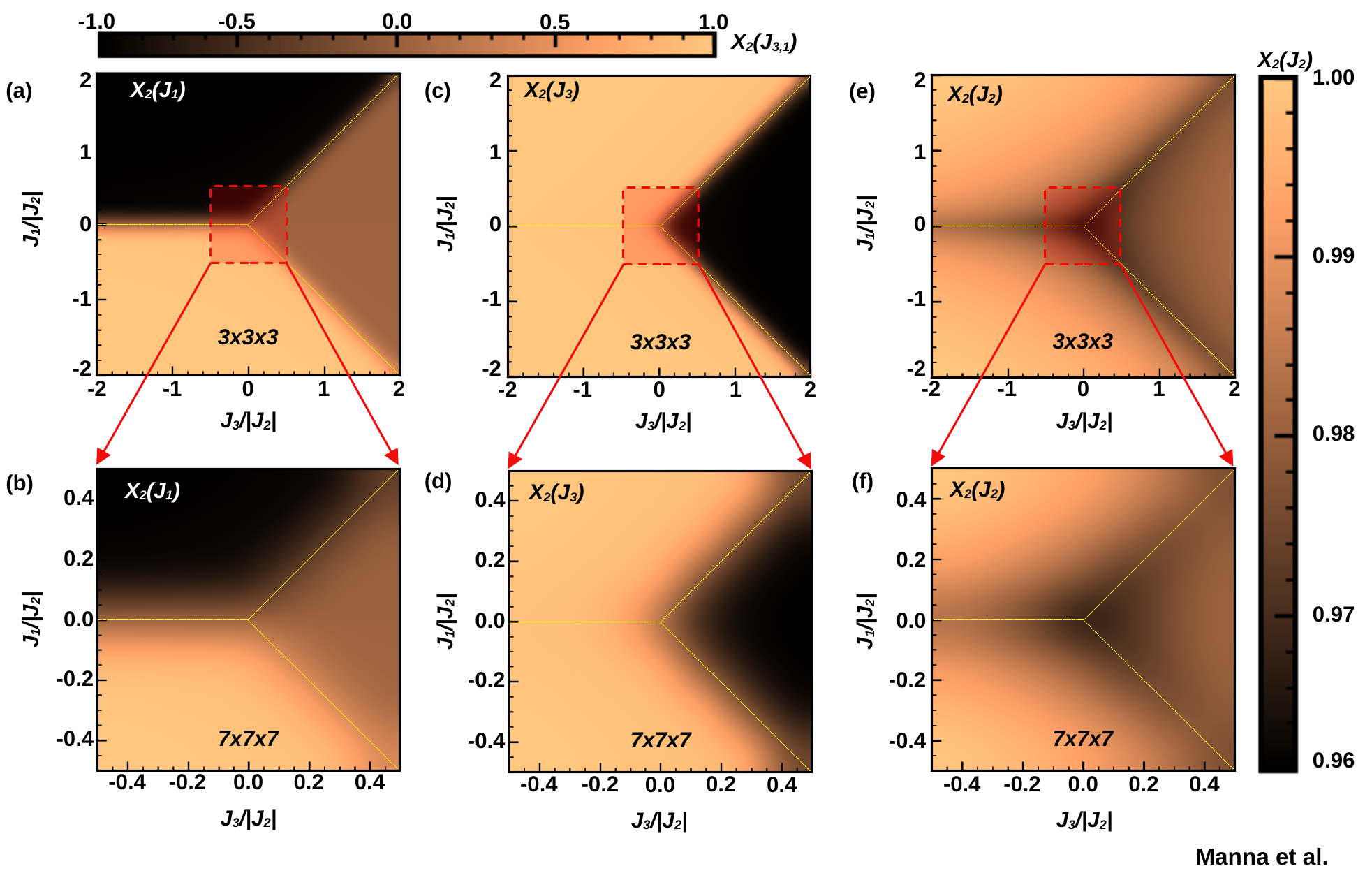}
\caption{Variation of the two-point spin--spin correlation function $X_2(J_i)$ ($i=1,2,3$) across the interaction parameter space obtained from Monte Carlo Simulations. The quantities $X_2(J_1)$, $X_2(J_2)$, and $X_2(J_3)$ denote correlations along the edges of the square units ($J_1$ bonds), along the bonds connecting the corners of adjacent square units ($J_2$ bonds), and along the diagonals of the square units ($J_3$ bonds), respectively. Panels (a), (c), and (e) present the correlation functions $X_2(J_1)$, $X_2(J_3)$, and $X_2(J_2)$, respectively, calculated for a lattice size of $3\times3\times3$ (216 spin sites) over a large parameter region $J_1/|J_2|, J_3/|J_2| \in [-2,2]$. Panels (b), (d), and (f) present the correlation functions $X_2(J_1)$, $X_2(J_3)$, and $X_2(J_2)$, respectively, for a larger lattice size of $7\times7\times7$ (2744 spin sites) over a selected parameter region $J_1/|J_2|, J_3/|J_2| \in [-0.5,0.5]$, highlighting the phase-boundary region around the bifurcation point.}
    \label{fig:Fig 3}
    
\end{figure*}

Fig. 2(a) summarizes the phase topology obtained from $\lambda_{\min}(\mathbf q)$. (i) An antiferromagnetic (AF) configuration is stabilized in which nearest neighbors within each square plaquette alternate in sign, whereas the corner-connecting bonds align in the same direction. This AF pattern is realized in both the frustrated and unfrustrated sectors and remains robust as long as $J_3>0$ (antiferromagnetic) does not exceed $J_1>0$ (antiferromagnetic) in magnitude (i.e., $J_3\lesssim J_1$). 
(ii) A ferromagnetic (FM) state is likewise remarkably stable: the system remains uniformly magnetized even when $J_3$ is antiferromagnetic, provided $|J_3|<|J_1|$ with $J_1<0$ (ferromagnetic). 
In both cases the dominant nearest-neighbor scale $|J_1|$ controls the selection between AF and FM order. The ferromagnetic (FM) phase is conventional, by contrast the antiferromagnetic (AF) regime is not a simple N\'eel state. Spins alternate in sign within each square plaquette, while the corner--to--corner links align ferromagnetically due to ferromagnetic inter--square coupling $J_2$. 
(iii) When $J_3$ is antiferromagnetic and becomes larger than $|J_1|$, the LT landscape enters a strongly frustrated regime in which conventional single-$q$ solutions violate the hard-spin constraint. Concretely, $\lambda_{\min}(\mathbf q)$ develops symmetry-related one-dimensional manifolds of minima confined to the $q_z=0$ plane, running along (in the first Brillouin zone) $q_x = \pm q_y$. Therefore, every mode with $\mathbf q=(q,q,0)$ and $\mathbf q=(q,-q,0)$ is degenerate. This places the LT spectrum in a broader class of soft-mode-manifold problems \cite{Bergman2007NatPhys,Gao2017NatPhys,Yao2021FrontPhys,Niggemann2019JPCM,GaoLineGraph2022PRL}. LT thus provides only a lower bound to the true ground-state energy in this domain. This magnetic structure suggests in-plane spirals uniform along $\hat z$; however, because the entire set of $(qq0)$ and $(q\bar q0)$ modes is degenerate, the physical ground state generally involves a continuous superposition of wave vectors on these lines rather than a single helix:
\begin{align}
\mathbf S_{\mu}(\mathbf r)
&=\sum_{\sigma=\pm 1}\int_{\mathcal C}\! dq\,\Big\{
   \mathbf A^{\mu}_\sigma(q)\cos\big[(q,\sigma q,0)\!\cdot\!\mathbf r+\phi^{\mu}_\sigma(q)\big]
   \nonumber\\
&\hspace{4.0em}
 +\;\mathbf B^{\mu}_\sigma(q)\sin\big[(q,\sigma q,0)\!\cdot\!\mathbf r+\psi^{\mu}_\sigma(q)\big]
\Big\},
\label{eq:superposition}
\end{align}
with amplitudes constrained by $|\mathbf S_\mu|=1$ [$\mu=1,2,\dots,8$], in the spirit of long-period modulated structures \cite{Izyumov1984SPU}. The resulting real-space texture need not resemble a simple helical pattern and may appear disordered in the $xy$ plane due to interference of many Fourier components. The functions $A^{\mu}_{\sigma}(\mathbf{q})$, $B^{\mu}_{\sigma}(\mathbf{q})$ are chosen in such a way so as to minimize the total free energy under the hard spin constraint. 

\begin{figure*}
     \centering
    \includegraphics[width=\textwidth]{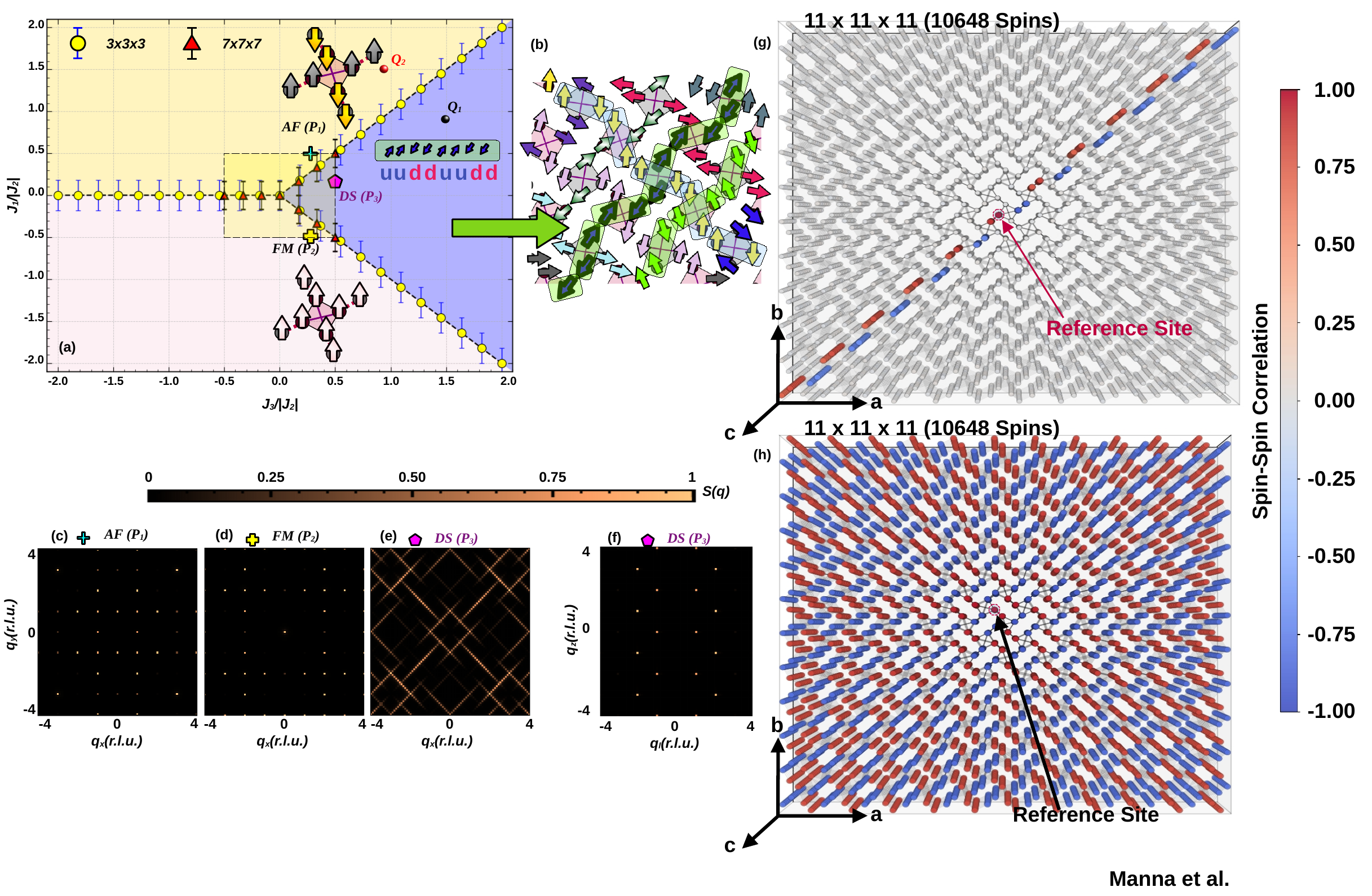}
\caption{\textbf{(a)} Low temperature ($T$=0.1$|J_2|$) magnetic phase diagram obtained from large-scale Monte Carlo simulations. Circular markers denote the phase-boundary points for the smaller lattice \(3\times 3\times 3\) (216 spin sites), and triangular markers denote those for the larger lattice \(7\times 7\times 7\) (2744 spin sites). The light-yellow and red regions indicate the antiferromagnetic (AF) and ferromagnetic (FM) phases, respectively, with representative spin configurations. The blue region corresponds to the DS phase. The three phase boundaries are \(J_{1}=-J_{3}\) (for \(J_{3}>0\)), \(J_{1}=J_{3}\) (for \(J_{3}>0\)), and \(J_{1}=0\) with \(J_{3}<0\). Three representative points $[P_1$ $(J_1/|J_2|=0.5, J_3/|J_2| =  0.3)$: AF$]$, $[P_2$ $(J_1/|J_2| = -0.5, J_3/|J_2|=0.3)$:  FM$]$ and $[P_3$ $(J_1/|J_2|=0.3, J_3/|J_2|=0.5)$: DS$]$  for which the corresponding reciprocal space plots of the static spin structure factor are displayed. 
\textbf{(b)} Schematic illustration of the DS phase, where magnetic order develops along the diagonal bonds and the corner-connecting links of the square units; however, correlations between these paths (highlighted in various colors) are highly suppressed. \textbf{(c)--(e)} Static spin structure factor for the Antiferromagnetic, Ferromagnetic and DS phases, respectively, at $q_z =0$ for a lattice of size 11$\times$11$\times$11 (10648 spins). \textbf{(f)} Static spin structure factor for the DS phase along $q_l$, $q_z$ space, $q_l$ being the reciprocal direction along the square diagonals. \textbf{(g), (h)} Spin-spin correlation maps in a \(11\times11\times11\) lattice (10648 spins) for the AF \((P_1)\) and DS \((P_3)\) states. One representative reference site is shown for each phase. The AF map confirms the long-range propagation of the plaquette-adapted staggered pattern, whereas the DS map shows that robust correlations propagate along the diagonal-corner-link trajectories associated with the \(uudd\) motif, with negligible spreading through the \(J_1\) edge direction.}

    \label{fig: Fig 4}
\end{figure*}
Further insights of the DS state has been obtained from the band structure (Momentum dependent eigen values $\lambda_{\mu}(q))$. The band structure is obtained from diagonalizing the interaction matrix along the trajectory $\Gamma\!-\!X\!-\!M\!-\!\Gamma\!-\!Z\!-\!R\!-\!A\!-\!Z$ [Fig. 2(b)]. Along the high-symmetry line $\Gamma\!-\!X\!-\!M\!-\!\Gamma\!-\!Z$, the bands exhibit the most pronounced dispersion, indicating strong momentum-space hybridization between the two sub-layers $z=L$ and $z=L+\frac{1}{2}$.  In the AF and FM regions, the energy surface in the \(\Gamma\!-\!X\) and \(\Gamma\!-\!M\) planes exhibits a clear global minimum at the zone center \(\Gamma\) [Fig.~2(f)]. Upon entering the frustrated sector with \(J_{3}/|J_{2}|>0\) and \(|J_{1}|>J_{3}\), this \(\Gamma\)-centered minimum progressively flattens, becoming noticeably shallower than in the corresponding unfrustrated cases with \(J_{3}<0\) [Fig.~2(g)]. Crossing the phase boundaries from the AF or FM regions into the DS regime, the energy surface evolves into a continuum of minima along the symmetry lines \(q_{x}=\pm q_{y}\) (in the  \(hk\)-plane) within the first Brillouin zone ($\Gamma-M$ in the reciprocal space) [Fig.~2(h)]. Along $Z\!\to\!R\!\to\!A \!\to\! Z$ path we observe that the eight LT modes collapse into three dispersing branches with enhanced degeneracies. This reduction follows from a decoupling of sub-layers $z=L$ and $z=L+\frac{1}{2}$, $L\in \mathbb{Z}$ [Fig. 1(c)].

\[
\gamma \equiv e^{i2\pi q_z},\qquad 1+\gamma=0 \;\;\Rightarrow\;\; \gamma=-1 \;\;\Rightarrow\;\; q_z=0.5 
\]
In a basis grouping the two layers (or two equivalent sublattice stacks), the Fourier-space exchange matrix takes the schematic block form
\begin{equation}
\underline{\underline{J}}(\mathbf q)=
\begin{pmatrix}
\mathcal J^{(1)}_{\mathrm{intra}}(q_x,q_y) & (1+\gamma)\,\mathcal T(q_x,q_y)\\[2pt]
(1+\gamma^\star)\,\mathcal T^\dagger(q_x,q_y) & \mathcal J^{(2)}_{\mathrm{intra}}(q_x,q_y)
\end{pmatrix}.
\label{eq:block}
\end{equation}
\begin{figure*}
    \centering
    \includegraphics[width=\linewidth]{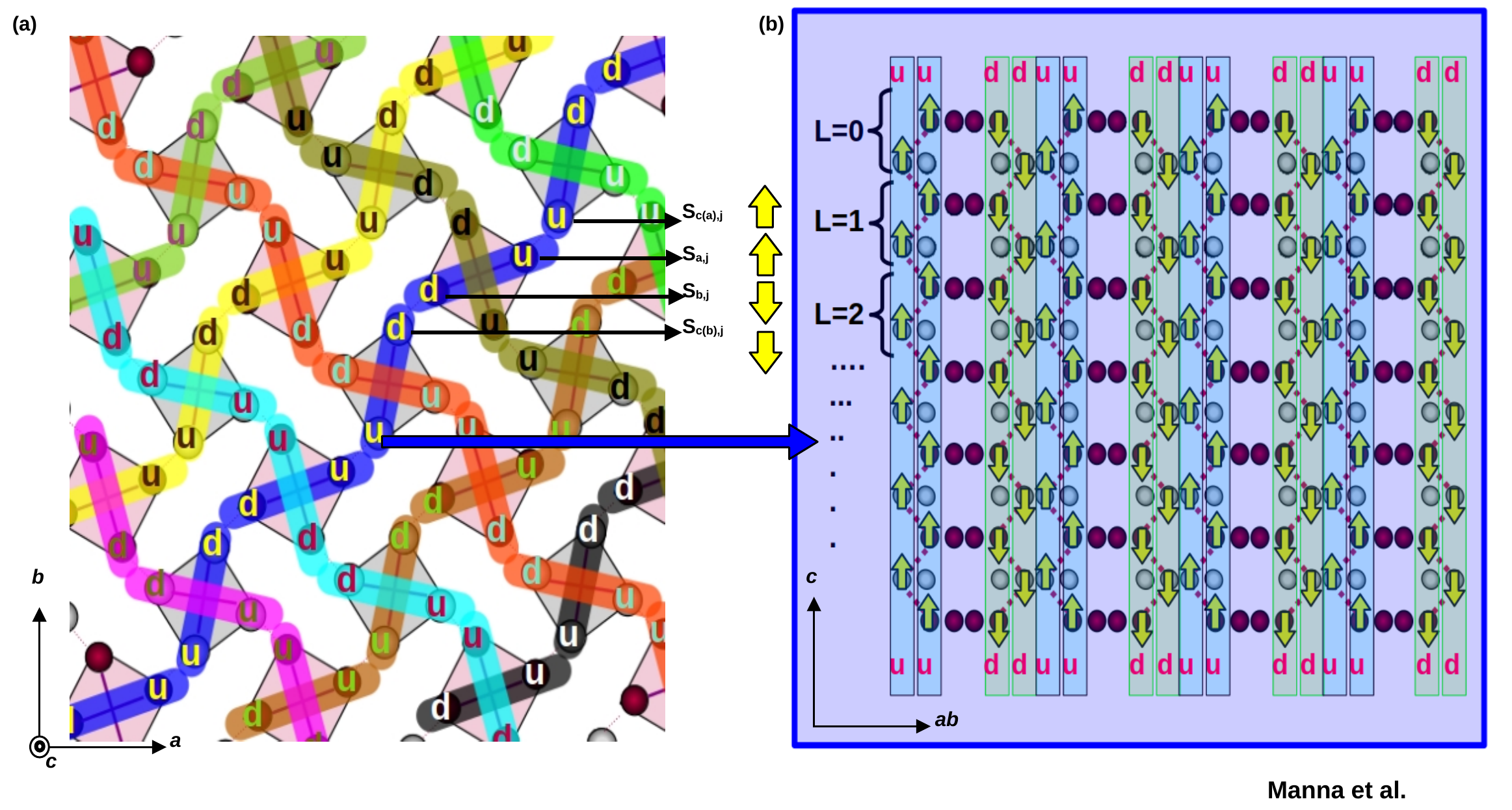}
    \caption{\textbf{(a)}~Top view (in the \textit{a–b} plane) of the spin configuration in the DS phase at low temperature $T=0.1|J_2|$. 
The colored diagonal trajectories represent paths, within which the magnetic moments exhibit a characteristic 
\textit{up–up–down–down} (\textit{uudd}) type of ordering, however, correlations between these paths are extremely small compared to the intra-path correlations. 
\textbf{(b)}~Sectional view along the $z$ direction corresponding to the blue trajectory in panel~(a). 
The \textit{uudd} sequence repeats ferromagnetically across successive layers, labeled by the layer index $L$ = 0, 1, 2, so on.
}
    \label{fig: Fig 5}
\end{figure*}
The block matrices (along the diagonal), $\mathcal J^{(1)}_{\mathrm{intra}}(q_x,q_y)$ and $\mathcal J^{(2)}_{\mathrm{intra}}(q_x,q_y)$ represent the contributions to the interaction matrix coming from the two sub-layers $z=L$ and $z=L+\frac{1}{2}$ [Fig. 1(c)]. On the other hand, the block matrices  $ (1+\gamma)\,\mathcal T(q_x,q_y)$ and its hermitian conjugate represent the contribution of the coupling between the two sub-layers to the interaction matrix. Precisely at $q_z=\frac{1}{2}$, the off-diagonal blocks vanish and Eq.~16 factorizes into two identical $4\times4$ blocks. The spectrum therefore comes in layer-doubled multiplets, yielding three distinct eigenvalue curves along $Z\!\to\!R\!\to\!A \!\to\! Z$ (with appropriate multiplicities) instead of eight unrelated branches.
\begin{figure*}
    \centering
    \includegraphics[width=0.85\linewidth]{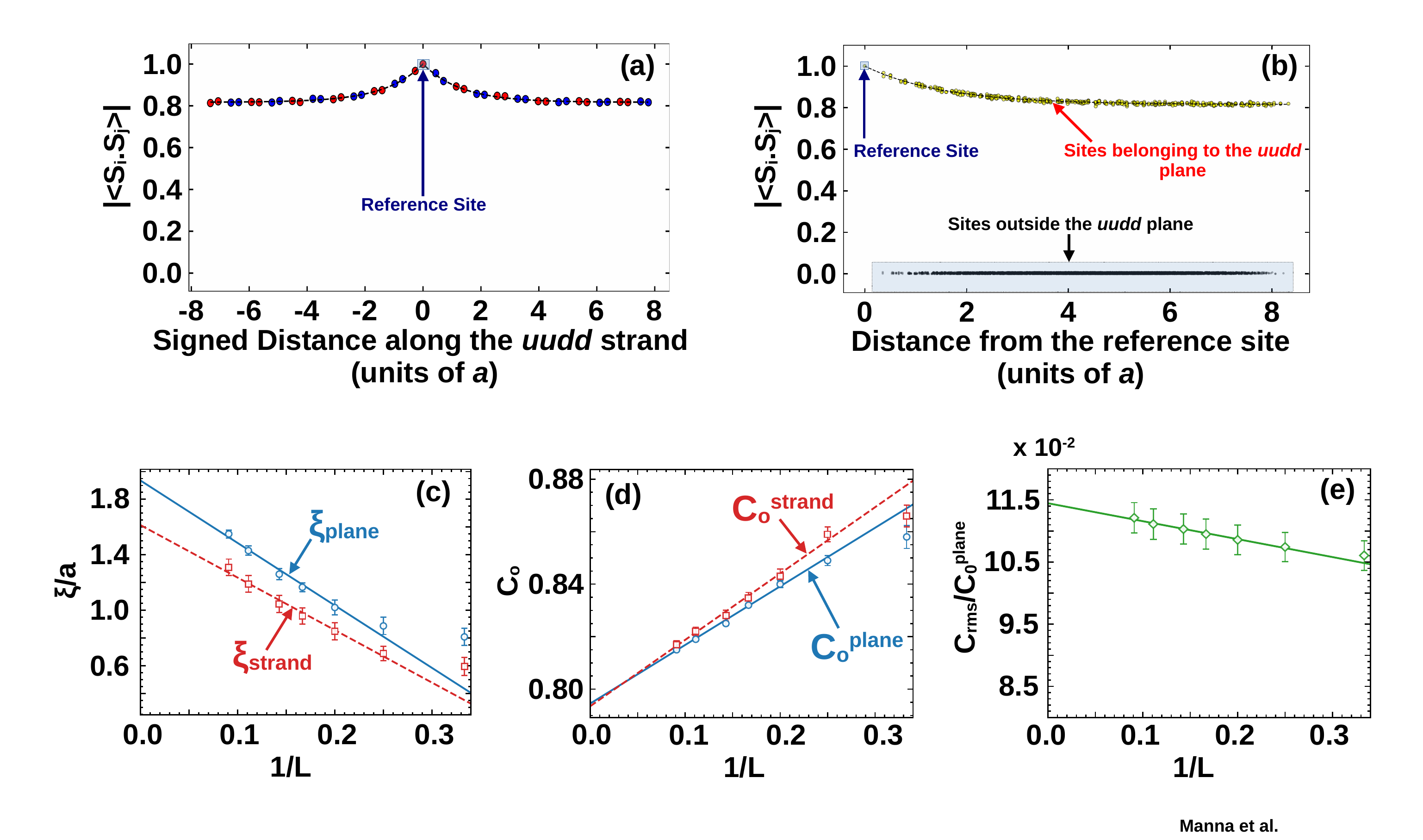}
    \caption{Finite Size study of the DS phase at the representative point $Q_1 (J_3/|J_2| = 1.5, \ J_1/|J_2| = 0.9)$: (a) Spin--spin correlations along the \textit{uudd} strand, plotted against the signed distance from a reference site. The red/blue markers represent  positive/negative correlation between a site on the strand and the reference site. (b) Magnitude of the spin--spin correlation, $\left|\langle\mathbf{S}{i}.\mathbf{S}{j}\rangle\right|$, as a function of the distance from a reference site for an $11\times11\times11$ lattice. (c) Correlation lengths along the \textit{uudd} strand, $\xi_{\mathrm{strand}}$, and within the \textit{uudd} plane, $\xi_{\mathrm{plane}}$, as a function of $1/L$. (d) Corresponding offset correlations, $C_{0}^{\mathrm{strand}}$ and $C_{0}^{\mathrm{plane}}$, obtained from the correlation-function fits, as a function of $1/L$. (e) Ratio $C_{\mathrm{rms}}/C_{0}^{\mathrm{plane}}$ as a function of $1/L$. The lines in panels (c)--(e) denote linear extrapolations toward the thermodynamic limit, $1/L\rightarrow0$. }
    \label{fig:6}
\end{figure*}

A second salient feature is the near-identity of the dispersions in the ferromagnetic and antiferromagnetic regions. On this network, flipping the sign of $J_1$ is equivalent to a sublattice gauge transformation, or, in reciprocal space, a shift of the ordering vector by a reciprocal-lattice half-vector $\mathbf Q$ (e.g.\ $\mathbf Q=(1/2,1/2,0)$):
\begin{equation}
\lambda_n(\mathbf q;\, J_1) \;=\; \lambda_n(\mathbf q+\mathbf Q;\, -J_1).
\label{eq:gauge}
\end{equation}

Because the plotted high-symmetry trajectory is mapped onto itself (up to relabeling) by $\mathbf q\!\mapsto\!\mathbf q{+}\mathbf Q$, the eigenvalue sets coincide, explaining why the band structures are same for two configurations with the same value of $J_3$ and same magnitude of $J_1$ but different signs [Fig.~2(c)--(h)]. Finally, the degeneracy associated with the frustrated sector is clearly visible along $M\!\to\!\Gamma$ [Fig.~2(e)]. There the lowest branch flattens and exhibits an extended softening consistent with a one-dimensional manifold of minima in the $q_z{=}\,0$ plane, specifically along $q_x=\pm q_y$. This LT signature is the reciprocal-space fingerprint of a degenerate soft-mode manifold, in which modes with $\mathbf q=(q,q,0)$ and $\mathbf q=(q,-q,0)$ are energetically equivalent and compete to form hard-spin-compatible multi-$\mathbf q$ superpositions in real space \cite{LuttingerTisza1946,LyonsKaplan1960,Bergman2007NatPhys,Niggemann2019JPCM,Yao2021FrontPhys}.  The LT analysis explains the remarkable robustness of AF and FM orders when $|J_3|<|J_1|$ (AF or FM depending on the sign of $J_1$), while identifying a strongly frustrated sector for AF $J_3>|J_1|$ in which $\lambda_{\min}(\mathbf q)$ exhibits line-like degeneracies. In the latter regime single-$q$ LT states are unphysical, and the ground state is expected to be a multi-mode, in-plane spiral superposition uniform along $z$. The combination of LT maps and Monte Carlo analysis provide a controlled route to determine magnetic behavior of such systems.

\subsection{Monte Carlo Analysis}
One needs detailed Monte Carlo simulations for an in-depth  investigation of the DS phase, validation of the LT calculations for the other phases (FM, AF) as well as study the properties of the spin-system as a function of temperatures and magnetic field.

The LT analysis provided the $T{=}0$ classical magnetic phase diagram [Fig. 2] in the $J_1$--$J_3$ plane and identifies a frustrated sector where single-$q$ LT states violate the hard-spin constraint. To validate the LT-predicted phase diagram and shed light on the degenerate spin state, we carried out large-scale classical Monte Carlo simulations, under both zero and applied magnetic fields (along the \textit{c} direction). This two-pronged approach: analytical LT at zero temperature together with Classical Monte Carlo analysis allows us to unambiguously determine the phase boundaries and, more importantly, to determine the true nature of the magnetic state in the frustrated portion of the parameter space at low temperatures.

We consider periodic lattices of $L_x\times L_y\times L_z$ conventional cells, each containing $N_s{=}8$ magnetic sites, so that the total number of spins is $N{=}N_s L_x L_y L_z$. Two global sweeps of the coupling space are carried out to balance resolution and system size: (i) a dense scan on a smaller lattice, $3{\times}3{\times}3$ ($N{=}216$ sites), using a $23{\times}23$ grid in $(J_1/|J_2|,J_3/|J_2|)$ such that both of these parameter ratios vary from $-2$ to $2$ [Figs.~3(a), (c), (e)]; and (ii) a confirmation scan on a larger lattice, $7{\times}7{\times}7$ ($N{=}2{,}744$ sites), using a $7{\times}7$ grid close to the bifurcation point $(J_1/|J_2|,J_3/|J_2|)=(0,0)$ with range $[-0.5,0.5]$ [Figs.~3(b), (d), (f)]. This strategy establishes the almost negligible size dependence of the phase boundaries while keeping the overall computational load manageable. The larger-lattice $7\times7\times7$ (2744 spin sites) scan focused around the bifurcation point of the phase boundary is particularly important because this is the region where the AF, FM, and DS tendencies compete most strongly; hence, any finite-size artefact would be most visible. The agreement between the two scans therefore provides a check that the nature of the boundary bifurcation and the emergence of the DS phase are intrinsic features of the system.

The MC phase diagram in Fig.~4(a) was obtained by assigning a phase label to each simulated point in the $(J_1/|J_2|, J_3/|J_2|)$ plane using a combined set of diagnostics. The ferromagnetic phase was identified by a finite total magnetization $M$ [Eq. (7)], together with positive correlations on all three exchange paths, i.e., $X_2(J_1) \to +1$, $X_2(J_2) \to +1$, and $X_2(J_3) \to +1$. The antiferromagnetic phase was identified by (i) a finite plaquette-adapted staggered magnetization $M_s$ [Eq.~(9)], (ii) vanishing total magnetization $M$, and (iii) the bond-correlation pattern $X_2(J_1) \to -1$, $X_2(J_2) \to +1$, and $X_2(J_3) \to +1$ [Eq.~(8)]. The DS phase was identified by (i) a finite four-spin order parameter $O_2$ [Eq.~(10)], (ii) negligible total magnetization $M$, and (iii) the characteristic correlation pattern $X_2(J_1)| \simeq 0$, $X_2(J_2) \to +1$, and $X_2(J_3) \to -1$. In addition to the nearest-neighbor bond-resolved correlations ($X_2$), we also checked long-range spin-spin correlations over the lattice size of $11\times11\times11$ [Fig.~4(g,h)] to ensure the phases. The phase boundaries in Fig.~4(a) were then obtained from the parameter points where these diagnostic parameters ($X_2$, $O_2$, $M$ and $M_s$) changed their characteristic behavior.

Given the frustrated character of the model, we employ parallel tempering \cite{HukushimaNemoto1996,EarlDeem2005PCCP,Katzgraber2006,Kofke2002,FerrenbergSwendsen1988PRL,FerrenbergSwendsen1989PRL} with 24 to 72 thermal replicas, to prevent the system from getting stuck into a local energy minima.  For each point $(J_1, J_3)$, we evaluate the first three bond correlations, which capture the ferromagnetic or antiferromagnetic tendencies within the square plaquettes and across the corner–connecting links [Fig.~3]. The correlations measured along the edges of the elementary square plaquettes reveal a clear trend. When the nearest–neighbor interaction is antiferromagnetic ($J_{1}>0$) and the diagonal interaction $J_{3}$ is either antiferromagnetic or ferromagnetic but satisfies $|J_{3}|<J_{1}$, the system enters the antiferromagnetic (AF) regime. In this sector, edge correlations approach $-1$ while diagonal correlations approach $+1$ [Figs.~3(a)–(d)], indicating alternating spins along plaquette edges and parallel alignment across diagonals. By contrast, the corner–to–corner links, directly influenced by ferromagnetic $J_{2}$, remain strongly positive [Figs.~3(e),(f)], evidencing parallel alignment across adjacent plaquettes. This combination defines an AF state that is not a simple N\'eel order. As a complementary momentum–space diagnostic, the static structure factor $S(\mathbf q)$ exhibits no intensity at $\Gamma$ in the AF phase despite a zone–center ordering vector, because the net moment per conventional cell vanishes [Fig.~4(c)]. In the ferromagnetic (FM) phase, $S(\mathbf q)$ peaks strongly at $\Gamma$ [Fig.~4(d)].

Between the boundaries, $J_{1}=J_{3}$ and $J_{1}=-J_{3}$ for $J_{3}>0$, the system realizes the DS phase. Its correlation pattern is distinctive: along square edges the correlations are extremely weak, along diagonals they are strongly negative, and along corner-to-corner links they remain nearly unity [Figs.~3(a)–(f)]. Real-space textures thus consist of periodic flips along diagonals with enforced parallel alignment on corner links [Fig. 4(b)]. The coexistence of highly suppressed edge correlations with robust diagonal and corner correlations indicates that the nominally three-dimensional lattice tends to develop an emergent quasi-two-dimensional spin-structure.

To verify the above interpretation, i.e. whether the local correlation rules inferred from the three exchange paths propagate over the full lattice, we have calculated full-lattice reference-site spin-spin correlation maps for representative low-temperature configurations on an $11\times11\times11$ lattice [Fig.~4(g)]. In the DS phase, the corresponding map shows finite correlations only along the diagonal bonds and corner-link trajectories. The correlations display the characteristic long-range $uudd$ pattern. Correlations outside the $uudd$ strand are highly suppressed. The top view (along the $c$ axis) of this spin-configuration is shown in  Fig. 5(a) showing strongly correlated $uudd$ strands. Because the exchange interaction $J_2$ is ferromagnetic, the $uudd$ strand extends uniformly along the $c$-direction forming a $uudd$ plane as shown in Fig. 5(b).
\begin{figure*}
    \centering
    \includegraphics[width=\linewidth]{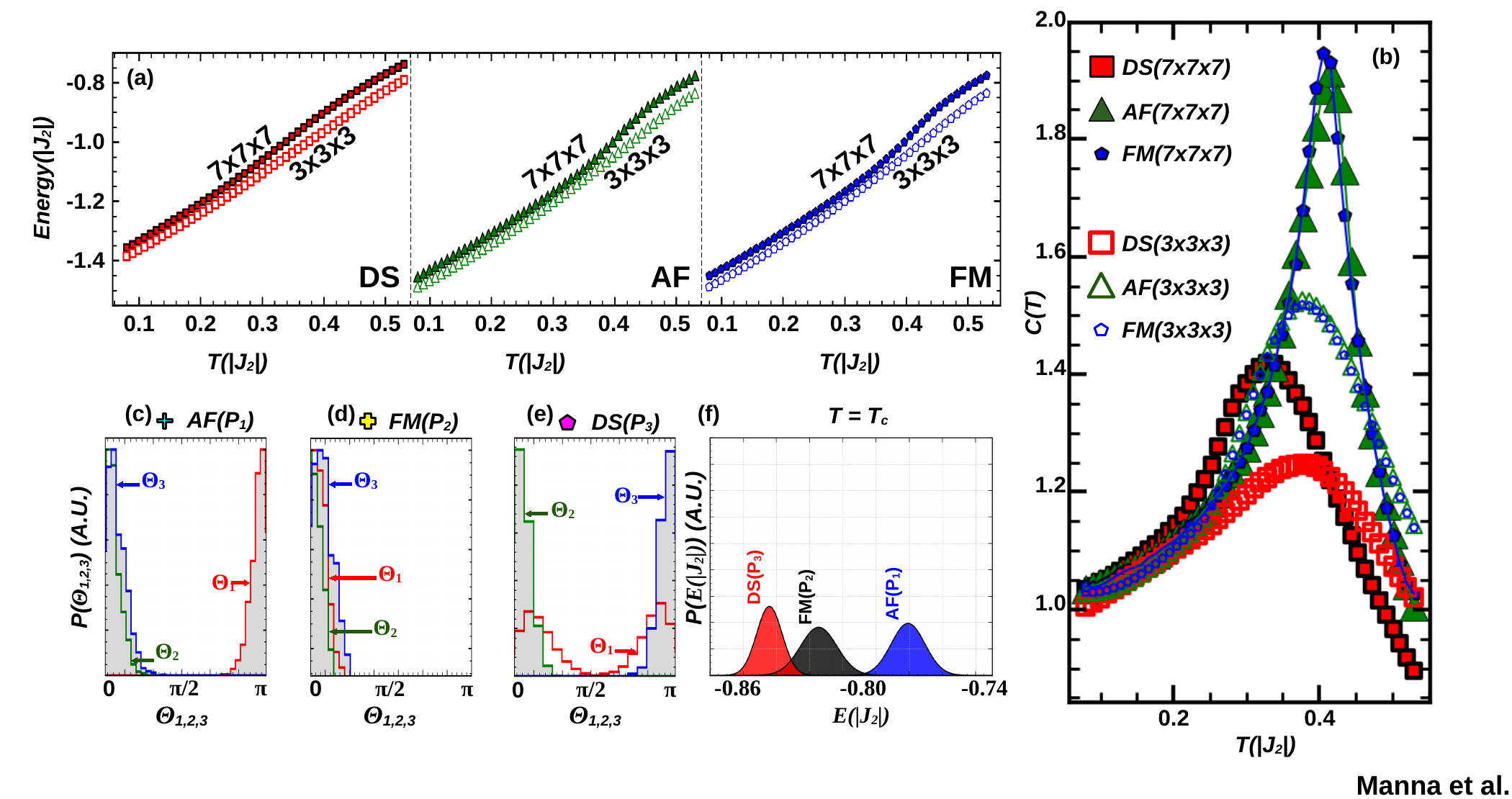}
    \caption{ Temperature variations of \textbf{(a)} energy and \textbf{(b)} specific heat for lattice sizes of $3\times3\times3$ (216 Spin sites) and $7\times 7\times 7$ (2744 Spin sites) for corresponding points taken are $P_1$ (AF), $P_2$ (FM) and $P_3$ (DS) [See Fig. 4(a)]. \textbf{(c), (d), (e)} Distribution of relative angle between spins along the sides of the square plaquettes $(\bm\Theta_1)$, diagonals of the square units $(\bm\Theta_3)$ and bonds connecting the corners of the square plaquettes $(\bm\Theta_2)$. \textbf{(f)} Distribution of energy at the critical temperature for the 7$\times$7$\times$7 lattice showing lack of bimodality. }
    \label{fig: Fig 7}
\end{figure*}

We further studied this interpretation by repeating the real-space correlation analysis at the representative point $Q_1(J_3/|J_2| = 1.5, J_1/|J_2| = 0.9)$ for several lattice sizes $(L = 3, 4, 5, 6, 7, 9\ \mathrm{and}\ 11)$ at $T = 0.1|J_2|$. The distances were generated based on the following ratio of lattice parameters: $a:b:c=2:2:1$. The spin-spin correlations were calculated for multiple reference sites in the lattice. The correlation function along a \textit{uudd} strand (for an $11 \times 11 \times 11$ lattice) is shown in Fig.~6(a), where the positive and negative values of distance represent the two opposite directions along the given strand with respect to the reference site. The full lattice correlation function, shown in Fig.~6(b) for an $11 \times 11 \times 11$ lattice, shows two branches having a strong spin-spin correlation with the \textit{uudd} plane and a weak spin-spin correlation outside the plane. The absolute value of correlation functions along the \textit{uudd} strand and within the \textit{uudd} plane were fitted to the equation $C(r)=A\exp(-r/\xi)+C_0$ to extract the corresponding correlation-decay lengths, $\xi_{\mathrm{strand}}$ and $\xi_{\mathrm{plane}}$, and the long-distance offsets, $C_0^{\mathrm{strand}}$ and $C_0^{\mathrm{plane}}$, for various lattice sizes. The fitted values of these quantities are shown against $1/L$ in Figs.~6(c) and 6(d) to reveal the finite size dependence. Finite-size linear extrapolations yield the values of correlation-decay lengths and the offsets as: $\xi_{\mathrm{strand}}/a=1.614\pm0.074$, $\xi_{\mathrm{plane}}/a=1.935\pm0.048$, $C_0^{\mathrm{strand}}=0.793\pm0.002$, and $C_0^{\mathrm{plane}}=0.794\pm0.001$, respectively. To measure the residual correlations outside the correlated plane, we additionally calculated $C_{\mathrm{rms}}$, the root-mean-square correlation between a reference spin site and all sites lying outside the \textit{uudd} plane. The ratio between $C_{\mathrm{rms}}$ and $C_0^{\mathrm{plane}}$ obtained for different lattice sizes is shown in Fig.~6(e). The thermodynamic-limit extrapolation of the ratio $\lim_{L\to\infty}C_{\mathrm{rms}}/C_0^{\mathrm{plane}}$ is determined to be $1.14(3)\times10^{-2}$ [Fig.~6(e)]. Thus the out-of-plane correlations is roughly $87$ times weaker than the in-plane correlations $C_0^{\mathrm{plane}}$ in the thermodynamic limit. This anisotropic pattern of correlations therefore elucidates an emergent quasi two-dimensional correlation structure within the underlying three-dimensional lattice of the DS phase.

\begin{figure*}
    \centering
    \includegraphics[width=\linewidth]{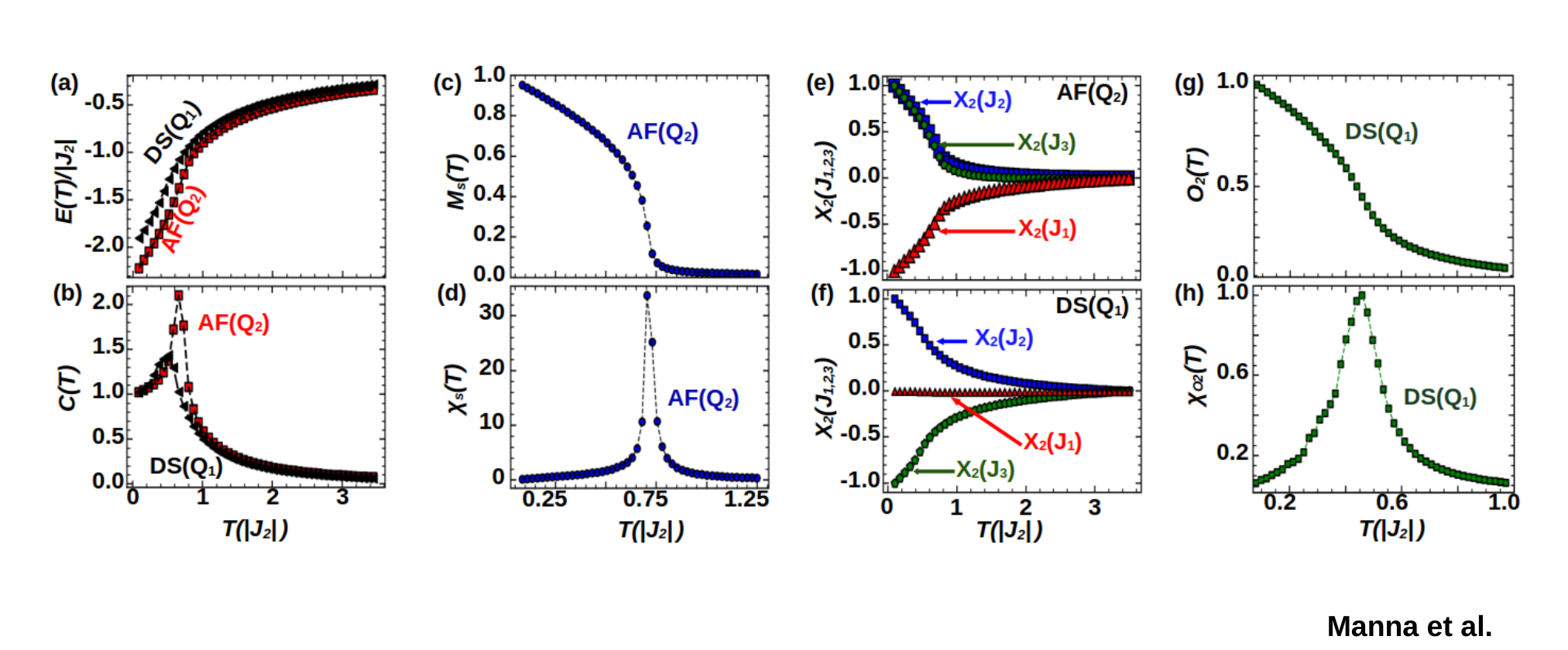}
    \caption{
Variation of thermodynamic observables with temperature at zero magnetic field calculated for representative points $\mathbf{Q_1}$ (DS: $(J_1=0.9,J_3=1.5)$), $\mathbf{Q_2}$ (AF: $(J_1=1.5,J_3=0.9)$) in Fig. 4(a) with a $11\times11\times11$ (10648 spins) lattice.
\textbf{(a), (b)} Energy \(E(T)\) and specific heat \(C(T)\) for both the DS and AF phases, respectively.
\textbf{(c), (d)} Plaquette-adapted staggered magnetization \(M_{s}\) (Eq.~9) and its susceptibility \(\chi_{m_{s}}\) (Eq.~11) for the AF phase.
\textbf{(e), (f)} Bond-resolved spin–spin correlations \(X_2(J_1)\), \(X_2(J_2)\), and \(X_2(J_3)\) along square edges, corner-connecting, and diagonal bonds for the AF and DS phases, respectively.
\textbf{(g), (h)} Order parameter \(O_2\) (Eq.~10) and its susceptibility \(\chi_{O_2}\) (Eq.~11) for the DS phase.}
    \label{fig:Fig 8}
\end{figure*}

In the reciprocal space, $S(\mathbf q)$ develops ridge-like maxima forming one-dimensional manifolds along $q_x\pm q_y=\text{const}$ in the $q_z=0$ plane [Fig.~4(e)], in line with the Luttinger-Tisza prediction of extended soft-mode lines [Fig.~2(h)]~\cite{Bergman2007NatPhys}. Long-range coherence along the $z$ axis within the DS manifold is confirmed by $S(q_z,q_l)$ [Fig.~4(f)], where $l$ tracks a diagonal trajectory. The associated four-spin order parameter $O_2$ is finite and its susceptibility $\chi_{_{O_2}}$ shows a pronounced peak at the transition. For comparison, the spin-spin correlation map for the AF phase is shown in Fig.~4(h). The correlation pattern generated from a single reference spin propagates coherently along all directions throughout the entire $11\times11\times11$ lattice, consistent with the plaquette-adapted staggered order and with the  bond correlations $X_2(J_1) \to -1$, $X_2(J_2) \to +1$, and $X_2(J_3) \to +1$.

Angle-resolved statistics provide an unbiased local probe that complements the bond-resolved correlations and the momentum-space structure factor discussed above [Figs.~7(c)-(e)]. We analyze the distributions $P(\bm\Theta_{1,2,3})$ of relative angles between neighboring spins along edges, diagonals, and corner links (see Eq.~6 for definitions). In the AF phase, the edge distribution $(\bm\Theta_1)$ is sharply skewed toward $180^\circ$, while the diagonal $(\bm\Theta_3)$ and corner-link $(\bm\Theta_2)$ distributions concentrate near $0^\circ$ [Fig.~7(c)], reflecting antiparallel alignment along plaquette edges together with parallel alignment across diagonals and between corner-connected plaquettes. This is accompanied by the onset of the plaquette-adapted staggered magnetization $\mathbf M_s$ (Eq.~9) and a pronounced maximum in its susceptibility $\chi_s$ (Eq.~11) at $T_c$ [Figs.~8(c),(d)], signaling the development of long-range AF order (discussed later in details). In the FM phase, all three angle distributions collapse toward $0^\circ$ with a finite width [Fig.~7(d)], consistent with uniform spin alignment on every bond type and the strong $\Gamma$-point intensity seen in $S(\mathbf q)$. The contrast with the AF case is thus immediate at the level of local angle statistics alone.

In contrast, the DS phase exhibits a richer pattern. The edge distribution $(\bm\Theta_1)$ becomes broad and \emph{bimodal}, with weight near both $0^\circ$ and $180^\circ$ and a continuous bridge between them, indicating that the system samples a wide range of edge orientations without settling into a unique ferromagnetic or antiferromagnetic arrangement. By contrast, the diagonal distribution $(\bm\Theta_3)$ is skewed toward $180^\circ$, reflecting robust anti-correlations across plaquette diagonals, while the corner-link distribution $(\bm\Theta_2)$ is skewed toward $0^\circ$, evidencing parallel alignment across the inter-plaquette links [Fig.~7(e)]. This triad, bimodal $(\bm\Theta_1)$, anti-aligned $(\bm\Theta_3)$, and aligned $(\bm\Theta_2)$, captures the location of the DS phase sandwiched between AF and FM in the low-temperature magnetic phase diagram and mirrors the quasi-two-dimensional network of the $uudd$ motif selected along diagonal trajectories.

Thermodynamic observables provide an independent probe of the ordering tendencies. 
Across all regions of the phase diagram, the specific heat exhibits a sharp anomaly at $T_c$, which we use to locate the transition [Fig. 7(b)]. The energy decreases monotonically with decrease in temperature with a higher slope near the transition temperature and is slightly larger in the DS phase as compared to AF or FM phases signaling enhanced frustration in the DS phase [Fig. 7(a)]. The energy probability distribution $P(E)$ in the vicinity of $T_c$ \cite{FerrenbergSwendsen1988PRL,FerrenbergSwendsen1989PRL,Binder1981ZPB} is unimodal for all the three magnetic phases [Figs.~7(f)], consistent with continuous (second-order) phase transitions. For the larger $11\times11\times11$ lattice at the representative points $Q_1$ and $Q_2$ [Fig. 4(a)], the zero-field energy and specific heat are shown in Figs.~8(a) and 8(b), respectively.

In the antiferromagnetic (AF) region, the plaquette-adapted staggered magnetization $M_s(T)$ [Eq.~(9)] onsets below $T_c$ [Fig. 8(c)], and its associated susceptibility [Eq.~(11)] develops a pronounced peak at the same temperature [Fig. 8(d)], signaling AF ordering. In the  degenerate spin state, the order parameter $O_2$ [Eq.~(10)] grows below $T_c$ and attains a large value at the lowest temperature [Fig.~8(g)], while its susceptibility $\chi_{_{O_2}}$ exhibits a peak at the transition temperature [Fig.~8(h)], providing a clear thermodynamic signature of the emergent \textit{uudd} order. Together, bond correlations, structure--factor maps, order parameters, angular--distribution analyses and full-lattice reference-site correlation maps  provide a consistent low-temperature characterization of the DS phase and connect the observed hard-spin correlation patterns to the LT topology.

To probe the magnetic response under an applied field, we consider two representative exchange parameter sets. In case (a), the parameters lie within the DS domain: \textbf{$\mathbf{Q_1}$} [Fig. 4(a)], and the simulations are performed on an $11\times 11\times 11$  lattice (10648 spin sites). In case (b), the parameters are chosen from the antiferromagnetic (AF) region: $\mathbf{Q_2}$ [Fig. 4(a)], again simulated on an $11\times 11\times 11$ lattice (10648 spin sites).

In the AF phase, we identify a critical curve in the field--temperature plane [Fig.~9(a)]: as the magnetic field increases, the corresponding critical temperature decreases in a non--linear manner. Below this curve, the AF character of the spin configuration is preserved. Specifically, within each square unit the spins alternate in sign, while along the corner--connecting bonds the spins align parallel to one another. Outside the critical curve at zero field (when $T>T_c(H=0)$), the system becomes paramagnetic. Similarly, when the field exceeds the critical field at low temperature, the system attains a field polarized state.

By contrast, the response in the DS regime is more nuanced [Fig.~9(b)]. We would like to recall, at zero field, this regime is characterized by dominant $uudd$-type correlations along diagonal–corner-link trajectories and strongly suppressed correlations through the square-edge channel. When a small field is applied along the $c$ direction at a low temperature, inter--layer correlations get enhanced. However, these correlations can be further suppressed by slightly increasing the temperature. Consequently, within the field--temperature ($H$--$T$) plane one can trace a trajectory along which the square edge spin--spin correlations remain substantially small. Below this trajectory, inter--layer correlations gradually build up and take negative values (negligible as compared to the intra--layer correlations); above it, the applied field promotes positive inter--layer correlations. Throughout this field--induced evolution, the order parameter $O_2$ remains large and positive until the state crosses outside the critical curve [Fig. 9(c)]. Thus, under a finite field and at temperatures below the corresponding transition temperature, the system in the DS domain can realize three distinct regimes: (i) a quasi--two--dimensional state with near zero edge correlations, (ii) a state with positively-correlated $uudd$ layers and (iii) a state with negatively correlated $uudd$ layers. Finally, once the critical curve is crossed, the system  becomes paramagnetic.

Within the corner-connected-square network, the DS state is characterized by line-like nearly flat LT minima running along symmetry-related directions in reciprocal space. This feature should be viewed in the broader context of soft-mode manifolds in frustrated magnets, including spiral surfaces on the diamond lattice, spiral contours/rings in honeycomb and square-lattice models, and related multiple-$\mathbf q$ instabilities \cite{Bergman2007NatPhys,Gao2017NatPhys,Gao2022PRL,Yao2021FrontPhys,Niggemann2019JPCM,GaoLineGraph2022PRL,PhysRevB.93.085132,Shimokawa2019PRL}. However, the present DS state is not a spiral spin liquid: the Monte Carlo simulations show a thermodynamically ordered multi-$\mathbf q$ state with a finite order parameter $O_2$, strong antiferromagnetic diagonal correlations, ferromagnetic corner-link correlations, and nearly vanishing edge correlations. Thus, the relevant comparison is not only to spiral-liquid phenomenology, but also to multiple-$\mathbf q$ ordered states \cite{Villain1980JPhys,Henley1989PRL,Okubo2012PRL,Hayami2016PRB,Hayami2024PRB,Gutzeit2022NatCommun}. Because a single-$\mathbf{q}$ configuration drawn from a line manifold cannot satisfy the hard-spin constraints on all sublattices,  The ultimate selection mechanism stabilizes a specific multi-$\mathbf{q}$ superposition, in close spirit to classic scenarios on other frustrated lattices, but here acting on a line rather than a ring or a surface of soft modes \cite{Villain1980JPhys,Henley1989PRL,Okubo2012PRL,PhysRevB.93.085132}. Taken together, these ingredients define a DS phase that is multi-$\mathbf{q}$ by necessity, plaquette-encoded in real space, and topologically distinct from previously studied spin-systems in momentum space\cite{Bergman2007NatPhys}. The DS state has sharp, testable signatures that differentiate it from known spiral liquids. In elastic neutron scattering it must yield ridge-like maxima confined to symmetry lines with dimensional reduction from a 3D lattice to statistically independent 2D layers \cite{Bergman2007NatPhys,Mulder2010PRB,Gao2022PRL} [Fig. 4(e), (f)]. In real space, the plaquette order parameter (Eq. 10) that captures the \textit{uudd} motif on diagonal–corner paths grows in tandem with the suppression of edge-bond correlations [Fig. 8(f)], providing a bond-resolved signature.
\begin{figure}
    \centering
    \includegraphics[width=\linewidth]{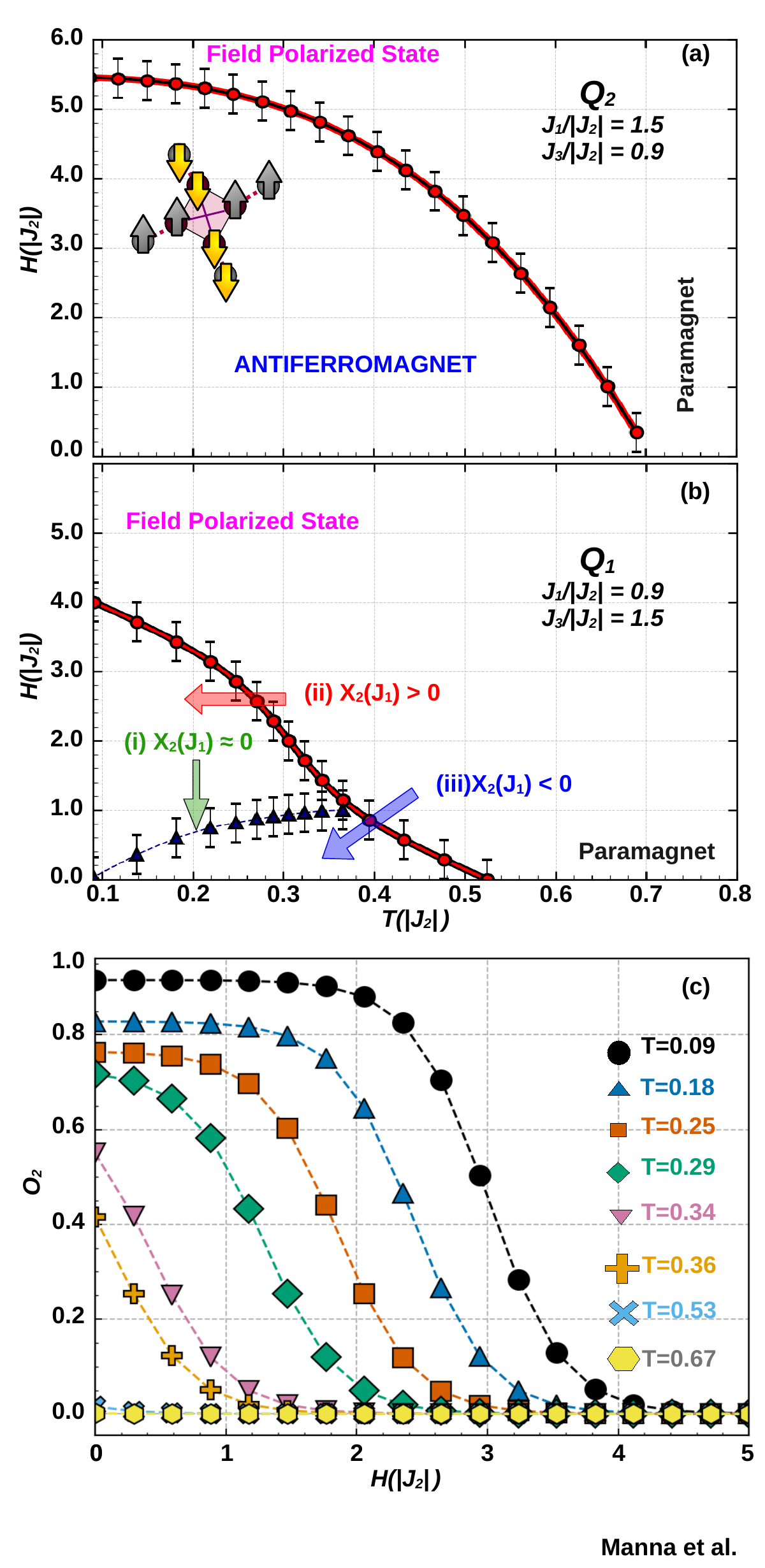}
    \caption{\textbf{(a)} \(H\text{--}T\) phase diagram (\(11\times 11\times 11\), 10648 spins) for the antiferromagnetic sector at \(Q_{2}\) [Fig. 4(a)].
    \textbf{(b)} Field–temperature \((H\text{--}T)\) phase diagram in the DS regime \((Q_{1})\). The bond correlation along \(J_{1}\) edges, \(X_{2}(J_{1})\), reveals a locus of highly-suppressed inter-layer correlations (navy blue curve). Below this curve, \(X_{2}(J_{1})<0\) (negatively correlated layers); above it, \(X_{2}(J_{1})>0\) (positively correlated layers). (c) The variation of the \(O_{2}\) order parameter versus magnetic field for several temperatures within the  DS regime at \(Q_{1}\) [Fig. 4(a)]. The system has \(11\times 11\times 11\) unit cells with 8 spins per cell (\(=10{,}648\) spins).}
    \label{fig:Fig 9}
\end{figure}

Our study captures the theoretical investigation of a Heisenberg hamiltonian embedded on the $X_{2}\mathrm{Re}_{3}\mathrm{Si}_{5}$ family (with rare-earth $X$) and thus provides a controlled, materials-motivated platform. Importantly, the present results establish a quantitative baseline obtained via LT and MC procedures on which more realistic ingredients can be added. In particular, single-ion anisotropies and long-range dipolar interactions associated with large rare-earth moments, deliberately omitted here, are natural next steps and are expected to enrich the phase diagram and lift residual degeneracies.

We would also like to point out that the nature of phase diagram with AF, FM and DS with $uudd $ ordering states [Fig. 4a] remains unchanged when the inter-plaquette interactions are weak. The FM and AF phase boundaries given by $|J_1| = |J_3|$ with $J_3 > 0$ in the phase diagram are sharply resolved within the LT framework and therefore remains unchanged irrespective of the value and sign of $J_2$. The physical spin configuration then emerges as a non-uniform superposition of equivalent $\textbf{q}$ modes in order to satisfy the hard-spin constraint. In agreement, even in the weak limit $\frac{J_3}{|J_2|} \sim 100$, only the AF, FM and DS with $uudd $ ordering states have been found from the Monte Carlo calculations. 

We also discuss the situation when the inter-plaquette interaction $(J_2)$ becomes antiferromagnetic for presently studied square-plaquette model. We find that the intra-plaquette spin-ordering is governed by the competition between $J_1$ and $J_3$, while $J_2$ primarily acts as a coupling that mediates correlations between the corners of neighboring plaquettes. Consequently, when $J_2$ is antiferromagnetic, the intra-plaquette ordering remains unaffected, whereas the spins connected by the inter-plaquette corner-sharing bonds become anti-aligned without any change in the phase boundaries (as shown in Fig. 4a).  For the AF phase, the spins continue to alternate within a plaquette and such AF plaquettes are now arranged antiparallel to each other, in contrast to the parallel alignments found earlier for a ferromagnetic $J_2$. Similarly, for the FM phase, all four spins of each plaquette remain parallel, whereas, neighboring plaquettes become anti-aligned, leading to an effective AF phase. For the degenerate spin state, the suppression of edge-correlations still persists which is consistent with the LT analysis, since the eigen-spectrum obtained from the diagonalization of the LT matrix is independent of the sign of $J_2$. However, the reversal of spins across the inter-plaquettes bonds leads to a change in the ordering from $uudd$ to $udud$ along each layer.

We view this work as a necessary first step towards a comprehensive theory of $X_{2}\mathrm{Re}_{3}\mathrm{Si}_{5}$, offering both a clean reference for future calculations (spin-wave, exact diagonalization, tensor networks) and a benchmark against which materials-specific effects can be systematically assessed.

\section{Summary and Outlook}

We have established a coherent picture of magnetic states for Heisenberg magnets on a three-dimensional network of corner-connected square plaquettes with competing edge ($J_1$), corner-link ($J_2$), and diagonal ($J_3$) exchanges. Our Luttinger-Tisza (LT) analysis, corroborated by large-scale hard-spin Monte Carlo (MC), identifies a zero-field topology comprising a robust antiferromagnet (AF), a ferromagnet (FM), and a highly frustrated degenerate state  with \textit{uudd} ordering.  In the DS regime, the lowest LT band develops line-like soft-mode manifolds, near-flat minima along $q_x=\pm q_y$  in the $hk$-plane and real-space bond fingerprints with extremely weak edge correlations, strongly antiferromagnetic diagonals, and ferromagnetic corner links leading to an emergent quasi-two dimensional $uudd$ spin ordering. A four-spin plaquette order parameter, $O_2$, cleanly diagnoses this state and aligns with modern diagnostics for multi-$\mathbf q$ competition \cite{Okubo2012PRL,Hayami2024PRB,Gutzeit2022NatCommun}.

Under applied field, we found that the AF sector exhibits the expected monotonic suppression of $T_c(H)$, whereas the DS state shows a quasi-two-dimensional response: a correlation-defined trajectory on which edge-bond correlations are highly suppressed, flanked by two quasi-2D states distinguished by the sign of interlayer correlations; notably, $O_2$ remains large until the state crosses into the paramagnet or a field polarized state. These field responses resonate with broader phenomenology of multi-$\mathbf q$ textures, conical/fan phases, and skyrmion crystals in frustrated centrosymmetric magnets   \cite{Nagamiya1967SSP,Okubo2012PRL,LeonovMostovoy2015NatCommun,Liu2024PRB,Hayami2024PRB}, suggesting that modest further ingredients (Single-ion anisotropy, small longer-range couplings) could realize tunable multi-$\mathbf q$ superpositions on the corner-connected square network.

We emphasize the following future directions in this field.
(i) Determination of spectral fingerprints and calculations related to dynamics \cite{Skubic2008JPCM,Evans2014JPCM, TothLake2015JPCM}. (ii) Materials mapping through first-principles workflows like DFT+$U$  \cite{Dudarev1998PRB,Liechtenstein1987JMMM, Aryasetiawan2004PRB, Setyawan2010CMS}. (iii) Experimental diagnostics. Diffuse neutron scattering and polarized-neutron analysis \cite{Squires2012,Blume1963PR,MoonRisteKoehler1969PR,Lynn2012MagNeutron} should reveal ridge-like intensity along $q_x=\pm q_y$ in the $hk$-plane in the DS sector, while the eight-site basis yields the usual Bragg selection rules separating AF (zero net moment per cell, no $\Gamma$ peak) and FM (strong $\Gamma$ intensity). Resonant elastic/inelastic X-ray scattering can provide element- and bond-selective sensitivity\cite{HillMcMorrow1996ActaA,Ament2011RMP}, allowing a direct comparison with the LT + MC results. Local probes such as $\mu$SR and NMR \cite{YaouancDeReotier2011,Slichter1990,Squires2012} should register the dimensional reduction in the DS manifold (layered slowing-down with weak interlayer locking) and track field-induced sign changes of interlayer correlations predicted here.

Therefore, corner-connected square plaquette spin-system hosts deceptively simple AF/FM orders alongside a rich DS manifold whose quasi-two-dimensional character survives modest fields and perturbations. By tying spectral topology (soft-mode lines) to bond-resolved observables and a robust multi-spin order parameter, our framework yields directly testable criteria for locating real materials in the phase diagram, steering them with field or strain along correlation-defined trajectories, and quantifying fluctuation-driven physics around the DS manifold with spin-wave, first-principles, and quantum many-body methods \cite{Okubo2012PRL,Hayami2017PRB,Hayami2024PRB,Bergman2007NatPhys}.

\end{document}